\documentclass[paper,11pt]{JHEP}
\usepackage[centertags]{amsmath}
\allowdisplaybreaks[1]
\usepackage{amsfonts} 
\usepackage{amssymb} 
\usepackage{amsthm}
\usepackage{graphicx}
\newcommand{\Z}{{\mathbb Z}}

\newcommand{\be}{\begin{equation}}
\newcommand{\eeq}{\end{equation}}
\newcommand{\bea}{\begin{eqnarray}}
\newcommand{\eea}{\end{eqnarray}}
\newcommand{\ba}{\begin{array}}
\newcommand{\ea}{\end{array}}

\newcommand{\eq}[1]{Eq.~(\ref{#1})}

\newcommand{\ii}{\mathrm{i}}
\newcommand{\ee}{\mathrm{e}}

\newcommand{\vev}[1]{\langle #1 \rangle}
\newcommand{\Tr}{\mathrm{Tr}\,}

\newcommand{\diag}{\mathrm{diag}\,}

\newcommand{\cF}{{\mathcal{F}}}

\newcommand{\cL}{{\mathcal{L}}}

\newcommand{\cN}{{\mathcal{N}}}

\newcommand{\cR}{{\mathcal{R}}}
\newcommand{\cS}{{\mathcal{S}}}
\newcommand{\cT}{{\mathcal{T}}}

\newcommand{\re}{\mbox{Re}\,}
\newcommand{\im}{\mbox{Im}\,}
\newcommand{\one}{{\rm 1\kern -.9mm l}}
\newcommand{\bone}{\mathbf{1}}
\newcommand{\Pf}{\mathrm{Pf}}
\newcommand{\ve}[1]{{\vec e}_{#1}}
\newdimen\tableauside\tableauside=1.0ex
\newdimen\tableaurule\tableaurule=0.4pt
\newdimen\tableaustep

\def\phantomhrule#1{\hbox{\vbox to0pt{\hrule height\tableaurule
width#1\vss}}}
\def\phantomvrule#1{\vbox{\hbox to0pt{\vrule width\tableaurule
height#1\hss}}}
\def\sqr{\vbox{%
 \phantomhrule\tableaustep
\hbox{\phantomvrule\tableaustep\kern\tableaustep\phantomvrule\tableaustep}%
 \hbox{\vbox{\phantomhrule\tableauside}\kern-\tableaurule}}}
\def\squares#1{\hbox{\count0=#1\noindent\loop\sqr
 \advance\count0 by-1 \ifnum\count0>0\repeat}}
\def\tableau#1{\vcenter{\offinterlineskip
 \tableaustep=\tableauside\advance\tableaustep by-\tableaurule
 \kern\normallineskip\hbox
   {\kern\normallineskip\vbox
     {\gettableau#1 0 }%
    \kern\normallineskip\kern\tableaurule}%
 \kern\normallineskip\kern\tableaurule}}
\def\gettableau#1 {\ifnum#1=0\let\next=\null\else
 \squares{#1}\let\next=\gettableau\fi\next}
\newcommand{\Yfund}{\tableau{1}}
\newcommand{\Ysymm}{\tableau{2}}
\newcommand{\Yasymm}{\tableau{1 1}}

\newcommand{\repst}[3]{(\mathbf{#1},\mathbf{#2},\mathbf{#3})}

\title{F-theoretic vs microscopic description of a conformal $\mathcal{N}=2$
SYM theory}
\author{\parbox{13.5cm}{Marco Bill\`o$^{1,2}$, Laurent Gallot$^3$, Alberto
Lerda$^4$
and Igor Pesando$^{1}$}
\\
~\\
~\\
$^1$Dipartimento di Fisica Teorica, Universit\`a di Torino\\
and I.N.F.N. - sezione di Torino \\
Via P. Giuria 1, I-10125 Torino, Italy\\
\vspace{0.15cm}
$^2$Kavli Institute for Theoretical Physics,
 University of California, Santa Barbara, \\
CA 93106-4030, USA\\
\vspace{0.15cm}
$^3$LAPTH, Universit\'e de Savoie, CNRS\\
9, Chemin de Bellevue\\
74941 Annecy le Vieux Cedex, France \\
\vspace{0.15cm}
$^4$Dipartimento di Scienze e Tecnologie Avanzate, Universit\`a del Piemonte
Orientale\\
and I.N.F.N. - Gruppo Collegato di Alessandria - sezione di Torino\\
Viale T. Michel  11, I-15121 Alessandria, Italy\\
\vspace{0.25cm}
\email{billo,lerda,ipesando@to.infn.it;
laurent.gallot@lapp.in2p3.fr}
}

\abstract{The F-theory background of four D7 branes in a type
I$^\prime$ orientifold was conjectured to be described by the Seiberg-Witten curve for
the superconformal SU(2) gauge theory with four
flavors. 
This relation was explained by considering in this background a probe D3 brane, which
supports this theory with SU(2) realized as Sp(1).
Here we explicitly compute the non-perturbative corrections to the D7/D3 system
in type I$^\prime$ due to D-instantons. This computation provides both the quartic
effective action on the D7 branes and the quadratic effective action on the D3
brane; the latter agrees with the F-theoretic prediction. The
action obtained in this way is related to the one derived
from the usual instanton calculus \`a la Nekrasov (or from its AGT realization in terms 
of Liouville conformal blocks) by means of a
non-perturbative redefinition of the coupling constant. 
We also point out an intriguing relation between the four-dimensional theory on the probe D3 brane 
with SO(8) flavor symmetry and the eight-dimensional dynamics on the
D7 branes. On the latter, SO(8) represents a gauge group and the
flavor masses correspond to the vacuum expectation values of an adjoint scalar field $m$: 
what we find is that the \emph{exact} effective coupling in four dimensions is obtained from its
\emph{perturbative} part by taking into account in its mass dependence
the full quantum dynamics of the field $m$ in eight dimensions.
}
\preprint{DFTT/7/2010\\LAPTH 023/10\\NSF-KITP-10-107 }

\begin{document}

\section{Introduction and motivations}
\label{sec:intro}
Phenomenologically viable string models
based on consistent D-brane
configurations have attracted a lot of attention in the last years 
\cite{Blumenhagen:2005mu}--\nocite{Blumenhagen:2006ci}\cite{Marchesano:2007de}.
In such constructions it is necessary to take into account possible
non-perturbative
corrections due
D-instantons and (wrapped) euclidean branes; for a review see, for example,
\cite{Blumenhagen:2009qh}. Some of these instantonic branes reproduce gauge
instantons
\cite{Witten:1995gx}--\nocite{Douglas:1995bn,Green:2000ke}\cite{Billo:2002hm}, 
other provide inherently stringy
(or ``exotic'') instanton effects; these latter can be responsible of important
terms in the effective action which would be perturbatively forbidden
\cite{Blumenhagen:2006xt}--\nocite{Ibanez:2006da}\cite{Florea:2006si}. 
Recently there has
been much
progress in the explicit computation of such contributions, both ordinary and
exotic, at least in supersymmetric cases (again, see \cite{Blumenhagen:2009qh}
and references therein). 

Another framework where phenomenological models with highly desirable features
can be set up, and where in particular Grand Unified Theories
occur naturally and consistently
\cite{Donagi:2008ca}--\nocite{Beasley:2008dc}\cite{Beasley:2008kw}, is
represented by F-theory
compactifications \cite{Vafa:1996xn} (for reviews see, for instance,
\cite{Denef:2008wq,Heckman:2010bq}).
F-theory gives a \emph{non-perturbative} geometric
description of type IIB backgrounds containing D7 branes and orientifold planes;
it somehow resums the non-perturbative corrections arising from certain
instantonic branes. Understanding in detail how this
resummation takes place would improve our knowledge of the relation between the
two descriptions. This could be useful for a better
comprehension of further non-perturbative effects in F-theory through a lift of
their type IIB counterparts, a subject that is recently%
\footnote{For an earlier discussion in an $\cN=2$ context see
\cite{Berglund:2005dm}.} receiving quite some attention
\cite{Blumenhagen:2010ja}--\nocite{Cvetic:2010rq,Alim:2009bx,Grimm:2009ef,
Grimm:2009sy}\cite{Jockers:2009ti}. 

In this paper we work out a simple, yet non-trivial, example where we are able
to compute the D-instanton effects in the IIB description and show that they
reconstruct the F-theory curve.  This example was considered by A.~Sen in
\cite{Sen:1996vd}, and is given by the compactification of F-theory on the
orbifold limit of an elliptically fibered K3 surface, for which the complex
structure modulus $\tau$ of the fiber is constant. This background was shown to
correspond to the so-called type I$^\prime$ theory, which is T-dual to type I
theory compactified on a 2-torus $T_2$, and thus possesses one O7 plane at each
of the four fixed points of $T_2$ with four D7 branes on top of it. Focusing on
the vicinity of one orientifold fixed-plane, and allowing the four D7
branes to move out of it, Sen conjectured that the corresponding F-theory
background should be described by the Seiberg-Witten (SW) curve
\cite{Seiberg:1994rs} for the $\mathcal N=2$ superconformal Yang-Mills theory
with gauge group SU(2) and $N_f=4$ flavors \cite{Seiberg:1994aj}. This relation
was later explained by T.~Banks, M.~Douglas and N.~Seiberg \cite{Banks:1996nj}
by considering a D3 brane in this background, which indeed supports such a
four-dimensional gauge theory on its world-volume, with SU$(2)$ realized as
Sp$(1)$. The SO$(8)$ flavor symmetry of the $N_f=4$ theory is nothing else but
the gauge group on the D7 branes. 

Here we explicitly compute the non-perturbative corrections to the D7/D3 system
in type I$^\prime$ due to D-instantons. This requires to identify the spectrum
of moduli, {\it i.e.} of excitations of the strings with at least one end-point
on the D-instantons, and the moduli action that arises from disks with at least
part of their boundary on a D-instanton, which was already discussed in
\cite{Gava:1999ky}. To obtain the non-perturbative effects it is necessary to
integrate over the moduli; we explicitly perform this integration by applying the
by-now standard techniques based on the BRST structure of the moduli spectrum
and action and its deformation by means of suitable RR backgrounds
\cite{Billo:2006jm}--\nocite{Billo:2009di,Fucito:2009rs}\cite{Billo':2010bd}.
This induces a complete localization of the integral, similarly to what happens
in supersymmetric instanton calculus in field theory
\cite{Nekrasov:2002qd}--\nocite{Flume:2002az,Bruzzo:2002xf,Nekrasov:2003af,
Nekrasov:2003rj}\cite{Bruzzo:2003rw}. These techniques have been recently
applied to similar brane set-ups, such as the D7 system in type I$^\prime$
\cite{Billo:2009di,Fucito:2009rs}, and the D7/D3 system on $T^4/\mathbb{Z}_2$
\cite{Billo':2010bd}.

D-instanton effects induce corrections to both the quartic
effective action on the D7 branes and the quadratic effective action on the
D3 brane. We use the prescription proposed in \cite{Billo':2010bd} to
disentangle the two contributions. The non-perturbative action on the D7's turns
out to be
exactly the same of the D7 system in type I$^\prime$ theory considered in 
\cite{Billo:2009di}. The
non-perturbative effective coupling on the D3 brane agrees with the
one extracted from the SW curve, that is with the F-theoretic prediction. 

One interesting question is the precise relation between the
eight-dimensional quartic effective action on the D7 branes and the F-theory
curve. This problem was already addressed in the past using the duality of
certain F-theory compactifications, including the one corresponding
to type I$^\prime$ theory, to heterotic models. Despite some interesting results
\cite{Lerche:1998nx,Lerche:1998gz},
this relation is not yet totally clear. Since in
our case the F-theory curve is nothing else but the SW curve encoding the
effective theory on the D3 probe, another way to state the above question is:
what is the relation between the non-perturbative effective actions on the D7
branes and on the D3 brane? More generally, how are the quantum dynamics on the
D7's and that on the D3 related to each other? 

We do not have a full answer to this question, but we uncover an intriguing
relation that goes as follows. On the D7's, the SO$(8)$ flavor
symmetry represents a gauge group and
the flavor masses $m_i$ correspond to the vacuum expectation values of an
adjoint 
scalar field $m(X)$: we find that the \emph{exact} effective four-dimensional coupling 
is obtained from its
\emph{perturbative} part by taking into account in its mass dependence
the eight-dimensional quantum dynamics of the field $m$, and in particular the so-called
``chiral ring'' formed by the correlators $\vev{\Tr m^{2l}}$.
In this way the F-theory geometry is related explicitly
to these eight-dimensional quantities.
It would be very interesting to investigate whether such sort of perturbative
propagation of the full quantum dynamics on a brane stack (in our case, the
D7's) to another one (in our case the D3 probe) takes place also in
other systems%
\footnote{Recently, D3 probes in F-theory have been considered in
\cite{Cvetic:2010rq,Heckman:2010fh}. In particular in \cite{Heckman:2010fh}
it is pointed out that the interplay between the eight-dimensional theory on the
D7 
and the four-dimensional theory on the probe, and the interpretation of
parameters in the
latter as adjoint fields on the former, plays a crucial r\^ole.}.

The theory that, in our example, lives on the D3 brane is
a four-dimensional conformal $\mathcal{N}=2$ theory. This class of theories
have recently attracted much attention in relation to the so-called AGT
conjecture put forward by L.~F.~Alday, D.~Gaiotto and Y.~Tachikawa
\cite{Alday:2009aq}. 
This conjecture relates the effective actions
obtained from usual instanton calculus \`a la Nekrasov \cite{Nekrasov:2002qd} 
to suitable correlators of the Liouville theory in two dimensions
\cite{Zamolodchikov:1985ie} (see
also
\cite{Marshakov:2009gs}--\nocite{Marshakov:2009kj}\cite{Poghossian:2009mk}). In
particular, the
non-perturbative action for the SU$(2)$ theory with $N_f=4$ can be
extracted from the 4-point functions on the sphere. It is interesting to compare
these results to what we get in the D7/D3 system in type I$^\prime$, where the
conformal theory is realized as an Sp$(1)$ gauge theory. It turns out that 
the effective action derived from the Nekrasov's prescription for the SU$(2)$,
$N_f=4$
case or, more efficiently, from its AGT realization in terms of Liouville
conformal blocks does not coincide, at first sight,
with our results, nor with the SW curve proposed in \cite{Seiberg:1994aj}.
There is a discrepancy already at the massless level: in this case from the SW
curve (and from our microscopic computation) we see that the tree-level
coupling receives no corrections. When computed following Nekrasov's
prescription for SU$(2)$, or by
the AGT method, instead, the coupling gets non-perturbatively modified%
\footnote{A discrepancy in this sense was already noticed at the two-instanton
level in \cite{Dorey:1996bf,Dorey:1996bn}, where the direct integration over the
moduli was performed without resorting to localization techniques.}. 
This suggests that a redefinition of the
coupling constant is needed in order to compare the two approaches; after
such a redefinition is performed, remarkably the two methods are reconciled
and the two results agree completely also in the massive case.

The paper is subdivided into several sections as we now describe.
In Section~\ref{sec:d3d7} we introduce the model and its F-theory description
through the SW curve; from this curve we extract the instanton expansion of the
effective coupling. In Section~\ref{sec:Dcorr} we describe the microscopic
computation of D-instanton contributions in our model. The resulting effective
action on the D7 branes is discussed in Section~\ref{sec:8d}, while in
Section~\ref{sec:4d} 
we write the D-instanton induced effective action on the D3 brane,
which is in full agreement with the F-theoretic description. In
Section~\ref{sec:intrel} 
we put forward our conjecture about the exact effective
coupling on the D3 being determined by its perturbative part plus the
eight-dimensional
dynamics of the mass parameters. Finally, in Section~\ref{sec:compare} we
discuss how our results compare to Nekrasov instanton calculus (or its AGT
realization) for  the SU$(2)$ theory with $N_f=4$, and in
Section~\ref{sec:concl} we present
our conclusions. In the appendices some
technical details and the extension of some results to asymptotically free
cases with $N_f<4$ are given.

\section{F-theory and the D7/D3 system in type I$^\prime$}
\label{sec:d3d7}
In F-theory compactifications over an elliptically fibered
manifold \cite{Vafa:1996xn}, the complex structure modulus 
$\tau$ of the fiber corresponds to the varying axio-dilaton 
profile of a suitable type IIB compactification on the base manifold.
In \cite{Sen:1996vd} A. Sen studied F-theory on an elliptically fibered K3
surface, which
is conjectured to be dual to heterotic string compactified on a two-dimensional
torus. 
He considered the particular case in which the K3 is at the orbifold limit in
moduli space where
it is described by the following curve in
Weierstrass form
\begin{equation}
 \label{fibtot}
 y^2 = x^3 -  \frac 14 G_2(z)\, x - \frac 14 G_3(z)~.
\end{equation}
Here $z$ is the coordinate on the base of the fibration and
\begin{equation}
 \label{GG}
  G_2(z) \propto Q^2(z)~,~~~
  G_3(z) \propto Q^3(z)~,~~~
 Q(z) = \prod_{I=1}^4 (z - f_I)~,
\end{equation}
with $f_I$ constants.
The absolute modular invariant of this curve
\begin{equation}
 \label{Jdef}
 J  = \frac{G_2^3}{G_2^3 - 27\, G_3^2}
\end{equation}
is $z$-\emph{independent}, and so is its complex structure modulus $\tau$
which can be determined from $J$ by inverting the relation
\begin{equation}
 \label{tauJ}
 J = \left(\frac{\vartheta_2^8(\tau) +
\vartheta_3^8(\tau) +
  \vartheta_4^8(\tau)}{24\, \eta^8(\tau)}\right)^3
\end{equation}
where the $\vartheta_a$'s are the Jacobi theta-functions and $\eta$ is the
Dedekind
function. 
By studying the metric on the base space, it can be seen that the latter has the
geometry
of a torus orbifold of the type $T_2/\mathbb{Z}_2$, with $\mathbb{Z}_2$ acting
as parity
reflection along $T_2$, in which the $f_I$'s appearing in (\ref{GG}) correspond
to the points of $T_2$ that are fixed under the $\mathbb{Z}_2$ action.

This specific F-theory background can be identified with the so-called
type I$^\prime$ theory, namely type IIB compactified on a torus $T_2$ and modded
out by
\begin{equation}
 \label{omedef}
  \Omega = \omega\, (-1)^{F_L}\, I_2~,
\end{equation}
where $\omega$ is the world-sheet parity reversal, $F_L$ is the left-moving
space-time fermion number and $I_2$ the inversion on $T_2$. 
This is the T-dual version of type I theory compactified 
on $T_2$, and possesses four O7
orientifold planes located at the points of $T_2$ that are fixed under $I_2$
(see Fig.~\ref{fig:d7t2}a). Each orientifold plane carries $(-4)$ units of 7-brane
charge, which need to be neutralized by putting 16 D7 branes transverse to
$T_2$.
If we place them in groups of 4 over each orientifold plane, the
tadpole cancellation becomes local and the axio-dilaton is constant over
$T_2$. From now on we take a local perspective and focus on one of the
orientifold fixed planes (say, the one at $z=f_1$) and its associated stack of 4
D7 branes.

\subsection{The local case}
\label{sec:local}
The action of the orientifold projection $\Omega$ is such that each group of 4
D7 branes supports an eight-dimensional theory with gauge group SO(8). Indeed,
the massless degrees of freedom of the 7/7 open strings build up an 
eight-dimensional chiral superfield in the adjoint representation of SO$(8)$, whose first few
terms are
\begin{equation}
M(X,\Theta) = m(X) + \sqrt{2}\,\Theta\,\sigma(X)
+\frac{1}{2}\, \Theta\gamma^{MN}\Theta\,f_{MN}(X) + \ldots~,
\label{Phi8}
\end{equation}
where $f_{MN}$ is the field-strength, $\sigma$ is the gaugino,
$m$ is a complex scalar, and $(X,\Theta)$ are the eight-dimensional super-coordinates. 
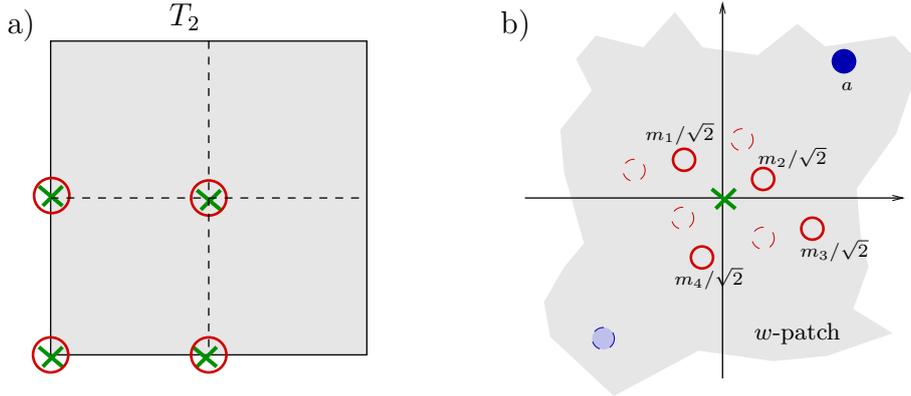
\begin{figure}[hbt]
 \begin{center}
\begin{picture}(0,0)%
\includegraphics{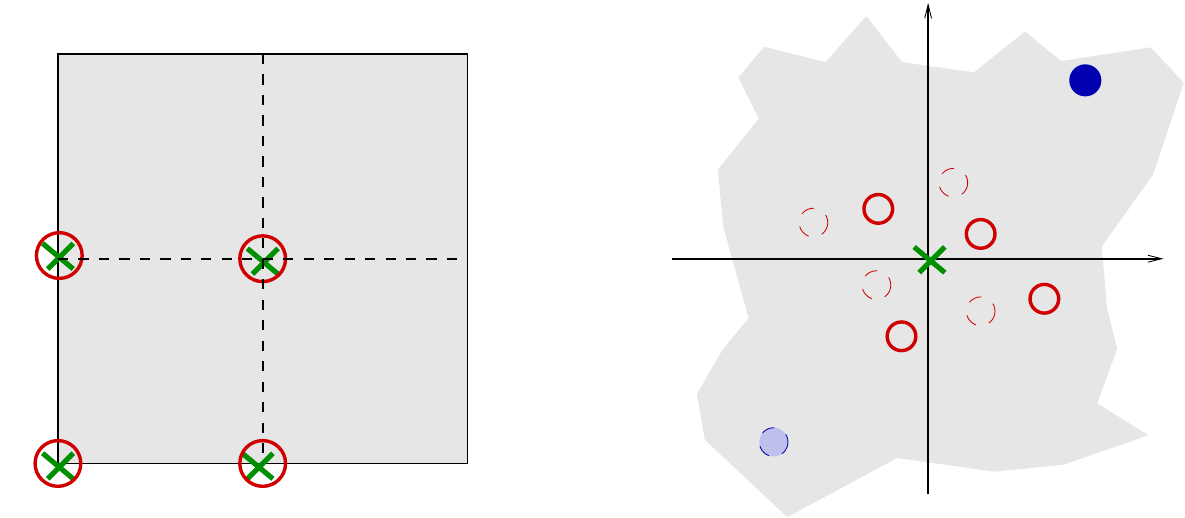}%
\end{picture}%
\setlength{\unitlength}{2155sp}%
\begingroup\makeatletter\ifx\SetFigFont\undefined%
\gdef\SetFigFont#1#2#3#4#5{%
  \reset@font\fontsize{#1}{#2pt}%
  \fontfamily{#3}\fontseries{#4}\fontshape{#5}%
  \selectfont}%
\fi\endgroup%
\begin{picture}(10408,4545)(391,-3684)
\put(2251,569){\makebox(0,0)[lb]{\smash{{\SetFigFont{12}{14.4}{\rmdefault}{\mddefault}{\updefault}$T_2$}}}}
\put(406,479){\makebox(0,0)[lb]{\smash{{\SetFigFont{12}{14.4}{\rmdefault}{\mddefault}{\updefault}a)}}}}
\put(7673,-743){\makebox(0,0)[lb]{\smash{{\SetFigFont{7}{8.4}{\rmdefault}{\mddefault}{\updefault}$m_1/\sqrt{2}$}}}}
\put(8953,-1023){\makebox(0,0)[lb]{\smash{{\SetFigFont{7}{8.4}{\rmdefault}{\mddefault}{\updefault}$m_2/\sqrt{2}$}}}}
\put(9433,-2083){\makebox(0,0)[lb]{\smash{{\SetFigFont{7}{8.4}{\rmdefault}{\mddefault}{\updefault}$m_3/\sqrt{2}$}}}}
\put(8003,-2403){\makebox(0,0)[lb]{\smash{{\SetFigFont{7}{8.4}{\rmdefault}{\mddefault}{\updefault}$m_4/\sqrt{2}$}}}}
\put(9903,-173){\makebox(0,0)[lb]{\smash{{\SetFigFont{7}{8.4}{\rmdefault}{\mddefault}{\updefault}$a$}}}}
\put(6026,479){\makebox(0,0)[lb]{\smash{{\SetFigFont{12}{14.4}{\rmdefault}{\mddefault}{\updefault}b)}}}}
\put(8911,-3031){\makebox(0,0)[lb]{\smash{{\SetFigFont{9}{10.8}{\rmdefault}{\mddefault}{\updefault}$w$-patch}}}}
\end{picture}%
 \end{center}
 \caption{~a) The initial set-up: 4 D7 branes (denoted by the red circles) are
 placed in the transverse torus $T_2$ at each of the O7 fixed planes (denoted by
 the green crosses).~~
 b) In the local limit near one of the fixed planes, the D7 brane displacements
 with respect to
 the original position are given by $m_i/\sqrt{2}$ (the image branes, denoted by
 dashed
 circles, sit at $-m_i/\sqrt{2}$). A D3 brane probe (depicted as a blue dot) is
 placed
 in $a$ (its image being in $-a$).}
 \label{fig:d7t2}
\end{figure}

If the D7 branes are moved away from the orientifold plane, {\it i.e.} if we
give a
diagonal vacuum expectation value to the scalar field $m$, the charges no longer cancel locally,
and correspondingly
the solution of the equation of motion for $\tau$ displays logarithmic
singularities at the orientifold and D7 brane locations.
Let us parametrize the region near the orientifold fixed point with a
coordinate%
\footnote{With respect to the global coordinate $z$ used above,
$w\propto (z - f_1)$.} $w$ and let the D7 branes and their images 
be located at $w=\pm m_i/\sqrt{2}$ with $i=1,\ldots,4$
(see Fig.~\ref{fig:d7t2}b). This corresponds to choose the
following vacuum expectation values%
\footnote{Notice that $m$ is a complex field in the adjoint of SO$(8)$, {\it
i.e.}
it is a complex antisymmetric matrix. If $m$ were real, its eigenvalues $m_i$
would be imaginary.\label{foot:m}} 
\begin{equation}
 \label{psitom}
  \vev{m} = \diag\big(m_1/\sqrt{2},\ldots,m_4/\sqrt{2},-m_1/\sqrt{2},
\ldots-m_4/\sqrt{2}\,\big)~.
\end{equation}
With this choice the axio-dilaton becomes
\begin{equation}
 \label{taubeh}
 \begin{aligned}
  {2\pi\ii}\,\tau(w) & = {2\pi\ii}\,\tau_0 + \left\{ \sum_{i=1}^4
\left[\ln\big(\,w -
  m_i/\sqrt{2}\,\big) + \ln\big(\,w + m_i/\sqrt{2}\,\big)\right] - 8\ln w
\right\}
 \end{aligned}
\end{equation}
where $\tau_0$ is the ``bare'' coupling.
This solution does not make sense everywhere, since
$\im\tau$ (proportional to the inverse string coupling) 
becomes negative close to the orientifold location ($w=0$). 
As is well known, this problem is cured 
by non-perturbative corrections. 

\subsection{F-theoretic description and D3 brane probes}
\label{subsec:FD3}
In \cite{Sen:1996vd} A. Sen proposed an F-theoretic interpretation of the local
set-up introduced above in which the exact axio-dilaton profile is given by
the complex structure modulus $\tau$ of the auxiliary Seiberg-Witten (SW) curve
that describes the quantum moduli space of the
$\mathcal{N}=2$ SYM theory in four dimensions with gauge group SU(2) and
$N_f=4$ fundamental flavors \cite{Seiberg:1994aj}.
The appearance of this seemingly unrelated four-dimensional gauge theory was explained in
\cite{Banks:1996nj} by considering D3 brane probes in the orientifold background
described in the previous section. As is well known, the orientifold projection
that leads to an orthogonal gauge group on the D7 branes gives rise to a symplectic gauge
group on the D3's. A single D3 and its image at an orientifold fixed point
support a $\mathrm{Sp}(1)\sim\mathrm{SU}(2)$ gauge group. The degrees of freedom of the
3/3 strings fill up a four-dimensional $\mathcal{N}=2$ chiral multiplet in the adjoint
representation
of Sp$(1)$, namely
\begin{equation}
 \Phi(x,\theta) = \phi(x) + \sqrt{2}\,\theta\,\Lambda(x)
+\frac{1}{2}\, \theta\sigma^{\mu\nu}\theta\,F_{\mu\nu}(x) + \ldots~,
\label{Phi4}
\end{equation}
where we denoted the four-dimensional super-coordinates as $(x,\theta)$. Notice that in this
orientifold
background the 3/3 spectrum contains also a neutral hypermultiplet%
\footnote{With $N$ D3 branes, this hypermultiplet transforms in the
antisymmetric representation of Sp$(N)$.}.

The open strings stretching between the D3's and the D7's contain precisely the
degrees of freedom of four fundamental hypermultiplets. From the D3 point of
view,
the SO(8) Chan-Paton group on the D7 branes represents the global flavor
symmetry group, while the D7 brane positions $m_i$ appear as masses for the
hypers,
since the 3/7 strings become stretched. It turns out that the neutral
hypermultiplet from the 3/3 sector is also completely decoupled from the
flavored matter, and thus it will be ignored in our subsequent analysis.

Moving the probe D3 brane away from the orientifold fixed point corresponds to
explore
the Coulomb moduli space of its world-volume theory. In the D3 brane action the
effective
coupling in front of the quadratic gauge Lagrangian for the chiral multiplet
in the Cartan direction is exactly given by the axio-dilaton
field, that is, we have the identification
\begin{equation}
\label{taugym}
\tau = \frac{\theta_{\rm YM}}{\pi} + \ii \frac{8\pi}{{g^2_{\rm YM}}}~,
\end{equation}
where $\theta_{\rm YM}$ is the theta-angle and $g_{\rm YM}$ the gauge coupling. 
This coupling receives perturbative corrections at the
one-loop level only, and for $N_f$ flavors it takes the form%
\footnote{In the $N_f=4$ case, the scale $\Lambda$ cancels out.}
\begin{equation}
 \label{tau1loop}
 {2\pi\ii}\,\tau_{\rm pert} = {2\pi\ii}\,\tau_0 + \Bigg\{\sum_{i=1}^{N_f}
 \Big[\ln\frac{a -m_i/\sqrt{2}}{\Lambda} + \ln \frac{a +
  m_i/\sqrt{2}}{\Lambda}\Big] 
 - 8 \ln \frac{a}{\Lambda} - 2 \ln 16\Bigg\}~,
\end{equation}
where $a$ parametrizes the vacuum expectation value of the adjoint Sp(1) scalar field, namely
\begin{equation}
 \label{defa} 
 \vev{\phi} = \diag (a,-a)~.
\end{equation}
For $N_f=4$ this expression is in agreement (a part from the finite
renormalization term
$2\ln 16$ on which we will come back later) with \eq{taubeh}, obtained by
solving
the field equations for the axio-dilaton sourced by the D7 branes and O7
planes. 
In this comparison we see that $a$ represents the position of the probe D3 brane
and
hence corresponds to the coordinate $w$.
For large $a$, the following expansion holds
\begin{equation}
 \label{taupertlf}
  \begin{aligned}
  {2\pi\ii}\,\tau_{\rm pert} & = {2\pi\ii}\,\tau_0 + \Bigg\{
  (2N_f - 8) \ln\frac{a}{\Lambda} - 2\ln 16 - \sum_{l=1}^\infty
  \frac{1}{2^l\,l}\, \frac{\sum_i m_i^{2l}}{a^{2l}}\Bigg\}\\
  & = {2\pi\ii}\,\tau_0 + \Bigg\{
  (2N_f - 8) \ln\frac{a}{\Lambda} - 2\ln 16 - \sum_{l=1}^\infty
  \frac{1}{2l}\, \frac{\Tr \vev{m}^{2l}}{a^{2l}}
  \Bigg\}~,
 \end{aligned} 
\end{equation}
where in the second line we used \eq{psitom}. This expansion will turn out to be
extremely
useful in the following.

As is well known, from this field theory point of view it
is possible to go beyond the perturbative results and derive
the \emph{exact} effective coupling on the moduli space of the D3 gauge theory 
from the appropriate SW curve. This coupling is then the exact 
axio-dilaton configuration for our set-up, {\it i.e.} it represents the
F-theoretic solution. 

\subsubsection{The SW curve for the D3 gauge theory}
\label{subsubsec:swc}

The SW curve for the SU(2) SYM theory with $N_f=4$ massive fundamental flavors 
was proposed in \cite{Seiberg:1994aj} to be given by a torus described by the
equation
\begin{equation}
 \label{swc}
 y^2 = P_3(x)~,
\end{equation}
where the cubic polynomial $P_3(x)$ is 
\begin{equation} 
 \label{swP}
  P_3(x) = W_1 W_2 W_3 + A \Big[W_1 T_1 (e_2 - e_3) + W_2 T_2 (e_3 - e_1) 
+ W_3 T_3 (e_1 - e_2)\Big]  - A^2 N~.
\end{equation}
Here, for $\ell=1,2,3$, we have introduced
\begin{equation}
 \label{swW}
  W_\ell = x - e_\ell\, \tilde u - e_\ell^2\, R~,
\end{equation}
while $R$, $T_\ell$ and $N$ are invariants of the flavor group SO$(8)$
that are, respectively, quadratic, quartic and sextic in the masses $m_i$ 
(see Appendix~\ref{app:flavinv} for details). The three quartic invariants
$T_\ell$ satisfy
the relation $T_1 + T_2 + T_3 = 0$, and hence we can use as independent
quantities
$T_1$ and the Pfaffian 
\begin{equation}
\label{pfaffdef}
\Pf m \equiv m_1 m_2 m_3 m_4~,
\end{equation}
since
\begin{equation}
 \label{Ttopf}
  T_2 = -\frac 12 \Big(T_1 + \Pf m\Big)~,~~~
  T_3 = -\frac 12 \Big(T_1 - \Pf m\Big)~.
\end{equation}
The quantities $e_\ell$ and $A$ are functions of the ``bare'' coupling $\tau_0$,
namely
\begin{equation}
 e_1 = \frac {\vartheta_3^4(\tau_0) + 
 \vartheta_4^4(\tau_0)}{3}~,~~~
 e_2 = \frac {-\vartheta_2^4(\tau_0)-
 \vartheta_3^4(\tau_0)}{3}~,~~~
 e_3 = \frac {\vartheta_2^4(\tau_0) -
 \vartheta_4^4(\tau_0)}{3}~,
\end{equation}
and
\begin{equation}
 \label{Adef}
 A = (e_1 - e_2)(e_2 - e_3)(e_3 - e_1) = 16\, \eta^{12}(\tau_0)~.
\end{equation}
Notice that all these functions can be expanded in power series with respect to
\begin{equation}
 \label{qdef}
  q = \ee^{\pi\ii\tau_0}~,
\end{equation}
which represents the weight of a gauge instanton configuration.
Finally, the parameter $\tilde u$ appearing in (\ref{swW}) is given by
\begin{equation}
 \label{utdef}
  \tilde u = u - \frac 12 e_1 R~,
\end{equation}
with $u$ being the SU$(2)$ invariant 
\begin{equation}
 \label{udef}
  u \equiv \vev{\Tr\phi^2}
\end{equation}
that parametrizes the Coulomb moduli space. 
In the semiclassical regime, {\it i.e.} when $u$ is large, we have
\begin{equation}
 \label{ua}
  u = \Tr \vev{\phi}^2 = 2 a^2~,
\end{equation}
where the second equality follows from (\ref{defa}).

Notice that in the massless case ($m_i=0$) the SW curve (\ref{swc}) reduces to
the
equation of a torus with
\begin{equation}
 \label{tefftau}
  \tau = \tau_0~.
\end{equation}
In the massive case, instead, the complex structure $\tau$ associated to the
curve
(\ref{swc}) differs from the ``bare'' coupling $\tau_0$. It
represents the exact coupling in the low-energy effective theory
on the D3 probe and hence the exact axio-dilaton background associated to the 
D7 brane configuration. We shall now extract from the SW curve the instanton
expansion of $\tau$ in the region of large $a$.

By a suitable shift in $x$, the SW curve (\ref{swc}) can be put in Weierstrass
form
\begin{equation}
 \label{wei}
  y^2 = x^3 - \frac{g_2}{4} x - \frac{g_3}{4}
\end{equation}
where the coefficients $g_2$ and $g_3$ can be determined explicitly in terms of
$u$, 
of the mass invariants and of $q$. The absolute modular invariant of the curve
\begin{equation}
 \label{jdef}
  j = \frac{g_2^3}{g_2^3 - 27\,g_3^2}
\end{equation}
has then an explicit expression in terms of the same parameters. This
expression is such that $j$ diverges as $q\to 0$; indeed
\begin{equation}
 \label{jdiv}
 j = \frac{1}{1728\, q^2}\frac{u^4}{\prod_i(u-m_i^2)} + O\big(q^{-1}\big)~.
\end{equation}
The modular invariant $j$ is related to the complex structure modulus 
of the torus by the expression given in (\ref{tauJ}), which we rewrite here
for convenience:
\begin{equation}
 \label{jtau}
  j = \left(\frac{\vartheta_2^8(\tau) +
\vartheta_3^8(\tau) +
  \vartheta_4^8(\tau)}{24 \,\eta^8(\tau)}\right)^3~.
\end{equation}
Inverting this relation yields the effective coupling as a function of $j$,
namely
\begin{equation}
 \label{tauj}
 {2\pi\ii}\,\tau = \ln\left(\frac{1}{1728}\,\frac{1}{j}\right)
  + \frac{31}{72}\,\frac{1}{j} + \frac{13157}{82944}\,\frac{1}{j^2} + \ldots~. 
\end{equation}
Then, using the expansion of $j$ given in (\ref{jdiv}), we can organize the
effective
coupling as a series in $q$ 
\begin{equation}
 \label{taue}
{2\pi\ii}\, \tau = {2\pi\ii}\,\tau_0 + \sum_k f_k(m,u)\, q^k
\end{equation}
in which each coefficient $f_k$ can in turn be expanded for large $u$.

The instanton expansion is then obtained by rewriting, in the semi-classical
regime, 
the modulus $u$ in terms of the vacuum expectation value $a$. To this
end, let us recall that the periods of the
holomorphic form $dx/y$ on the Weierstrass torus have an explicit expression in
terms of $j$. In the region where $j$ is large, one period has the following
closed-form expression
\begin{equation}
 \label{ome1def}
 \omega_1 = (48\, g_2)^{-\frac 14} F(\frac{1}{12},\frac{5}{12},1;\frac 1j)~,
\end{equation}
in terms of the hypergeometric function $F$,
and is related to the period $a$ of the SW meromorphic form by
\begin{equation}
 \label{ometoa}
  \frac{\partial a}{\partial u} = \omega_1~.
\end{equation}
Indeed, it is possible to expand $\omega_1$, and hence $a$, as a double series
in $q$ and in $1/u$, obtaining 
\begin{equation}
 \label{aexp}
  a(u) = \sqrt{\frac{u}{2}} \left(1 - 4\,\frac{\Pf m}{u^2}q + \ldots\right)~.
\end{equation}
Inverting this expansion, one gets the expression of $u(a)$ as a sum of
instanton contributions, valid for large $a$, namely
\begin{equation}
 \label{uaexp}
  u(a) = 2 a^2 + 4\, \frac{\Pf m}{a^2} q + O(q^2)
\end{equation}
in which the leading term is in agreement with the classical result (\ref{ua}).
Inserting this expansion into (\ref{taue}), we obtain%
\footnote{The expansions are rather cumbersome, so we limit ourselves to order
$q^3$ and to order $1/a^8$. This already provides a very strong check
against the result of the explicit instanton computations presented in the
following sections.} 
\begin{eqnarray}
  2\pi\ii\tau & =& 2\pi\ii\tau_0 -\frac{R}{a^2} + 
  \frac{1}{a^4} \left[\frac{-R^2 + 6 T_1}{4} + 12\, \Pf m \, q 
  + 6 (R^2 + 6 T_1)\, q^2 + 48 \,\Pf m~ q^3 + \ldots\right] \nonumber\\
  &&~~ + \frac{1}{a^6} \left[\frac{- 6 N - R^3 + 15 R T_1}{12} 
  - 60 (2N + R T_1) \, q^2 - 320 \,\Pf m\, R\, q^3 + \ldots  \right]\nonumber\\
  &&~~ + \frac{1}{a^8} \left[\frac{2 (\Pf m)^2 - 16 N R - R^4 + 28 R^2
  T_1 - 36 T_1^2}{32} + \frac{105}{2} (\Pf m)^2\, q^2\right.\nonumber \\
  && \left.\phantom{~~~~+ \frac{1}{a^6} \big[}+ 280\, \Pf m\, (R^2 +
  6 T_1)\, q^3 + \ldots\right]+~\ldots~.
\label{tauswe} 
\end{eqnarray}
Using the relations (\ref{Rdef}) and (\ref{combinv}) given in
Appendix~\ref{app:flavinv},
we can rewrite this expression as follows
\begin{eqnarray}
  2\pi\ii\tau & =& 2\pi\ii\tau_0 - \frac 12 \frac{\sum_i m_i^2}{a^2} + 
  \frac{1}{a^4} \Bigg[\!-\frac 18 \sum_i m_i^4 + 12\, \Pf m \,q 
  + 6 \!\sum_{i<j} m_i^2 m_j^2 ~q^2 + 48 \,\Pf m\,q^3 + \ldots\Bigg] \nonumber\\
  && + \frac{1}{a^6} \Bigg[\!- \frac{1}{24} \sum_i m_i^6 
  - 30 \sum_{i<j<k}m_i^2 m_j^2 m_k^2 ~ q^2 - 160\, \Pf m\, \sum_i m_i^2~ q^3 +
  \ldots \Bigg]\nonumber\\
  &&+ \frac{1}{a^8} \Bigg[\!-\frac{1}{64} \sum_i m_i^8 + 
  \frac{105}{2}\, (\Pf m)^2~ q^2
  + 280\, \Pf m\, \sum_{i<j} m_i^2 m_j^2
  ~ q^3 + \ldots\Bigg]+~\ldots~.
\label{tauswe2}
\end{eqnarray}
The effective coupling $\tau$ encoded in the SW curve contains a
perturbative part in perfect agreement%
\footnote{A part from the finite renormalization constant $2 \ln 16$.}
with \eq{taupertlf} for $N_f=4$ plus a series of non-perturbative corrections
$\delta_l$, and thus takes the form
\begin{equation}
 \label{tauswc}
 {2\pi\ii}\, \tau = {2\pi\ii}\,\tau_0 - \sum_{l=1}^\infty
  \frac{1}{2l} \,\frac{\Tr \vev{m}^{2l} + \delta_l}{a^{2l}}~, 
\end{equation}
where the explicit expressions for $\delta_l$ can be deduced from \eq{tauswe2}.
In particular, the first few corrections are
\begin{equation}
 \begin{aligned}
  \delta_1&=0~,\\
\delta_2&=- 48\, \Pf m ~q -24\sum_{i<j} m_i^2 m_j^2 ~ q^2 -192\, \Pf m~ q^3 +
\ldots~,\\
\delta_3&= 180\sum_{i<j<k} m_i^2 m_j^2
 m_k^2 ~q^2 + 960\, \Pf m\, \sum_i m_i^2~ q^3 + \ldots ~,\\
\delta_4&=- 420\, (\Pf m)^2 ~q^2 - 2240\,\Pf m\,
 \sum_{i<j}m_i^2 m_j^2~ q^3 + \ldots ~.
 \end{aligned}
\label{deltal}
\end{equation}
In the following we will provide a microscopic derivation of this result by
introducing
stacks of D-instantons in our orientifold model.

\section{D-instanton corrections}
\label{sec:Dcorr}
We now discuss the non-perturbative effects obtained by adding $k$ D-instantons to
the D3/D7 brane system described so far%
\footnote{According to the terminology used in the previous section this actually corresponds to
adding $k$ half-D(--1) branes plus their orientifold images.}. The D(--1)'s describe ordinary
gauge instantons in the four-dimensional $\mathcal N=2$ $\mathrm{Sp}(1)$ 
gauge theory with $N_f=4$ realized on the world-volume of 
the D3 branes, while they represent exotic instantons with respect to 
the eight-dimensional $\mathcal N=2$ $\mathrm{SO}(8)$ gauge theory 
realized on the D7 branes, as discussed in \cite{Billo':2009gc,Billo:2009di,Fucito:2009rs}.
In the following we describe the moduli spectrum and its BRST structure and
compute the D-instanton
partition function using localization methods.

\subsection{Instanton moduli spectrum}
\label{subsec:spectrum}

The open strings with at least one end-point 
on the D(--1)'s account for the instanton collective coordinates (or moduli)
and can be distinguished into three classes corresponding to open
strings with both end-points on the D-instantons (the (--1)/(--1) sector), and
to open strings with one end-point on the D-instantons and the other 
either on the D3 branes or on the D7 branes (the (--1)/3 sector and the (--1)/7 sector respectively). 
This system has already been considered and described in \cite{Gava:1999ky}, but for 
completeness we now briefly recall its main features, taking advantage of the analysis presented
in \cite{Billo':2010bd} for a very similar set-up.

\paragraph{(--1)/(--1) sector} The string excitations in this sector carry
Chan-Paton (CP)
factors that are $(k\times k)$ matrices in a definite representation of $\mathrm{SO}(k)$ 
selected by the orientifold projection $\Omega$.
They also organize in representations of the Lorentz
group which in our case is
\begin{equation}
 \mathrm{SO}(4) \times \widehat{\mathrm{SO}}(4) \times \mathrm{SO}(2) ~\simeq ~
\mathrm{SU}(2)_+\times \mathrm{SU}(2)_-\times
\widehat{\mathrm{SU}}(2)_+ \times \widehat{\mathrm{SU}}(2)_-
\times \mathrm{U}(1)
\label{lorentz}
\end{equation}
as dictated by the presence of the D7 and D3 branes. 

In the NS sector the two bosonic moduli along the D7 transverse directions, with ADHM dimension
of (length)$^{-1}$, are odd under 
$\Omega$ and thus transform in the anti-symmetric representation
of $\mathrm{SO}(k)$. 
They can be organized in complex conjugate moduli $\chi$ and $\bar\chi$
that are singlets of the four SU(2)'s in (\ref{lorentz}) and carry
charge $+1$ and $-1$ under the $\mathrm{U}(1)$ factor. The 
eight bosonic moduli along the D7 longitudinal directions, with ADHM dimension of
(length)$^{+1}$, are instead even under
$\Omega$ and hence transform in the symmetric representation 
of $\mathrm{SO}(k)$.
They form two vectors for each of the two $\mathrm{SO}(4)$ factors in (\ref{lorentz}) 
that we denote, respectively, $a^\mu$ ($\mu=0,\ldots,3$) and $a^m$ ($m=4,\ldots,7$). The diagonal
components $\Tr a^\mu$
define the (center of mass) position of the D-instantons
in the four-dimensional space that is longitudinal to both the D7's and
the D3's, and thus they can be identified with the four-dimensional space-time coordinates $x^\mu$. 
The other four diagonal components $\Tr a^m$
represent the position in the remaining four longitudinal directions of the D7 branes,
and together with the $x$'s they can be identified with the eight-dimensional coordinates $X$ used in
the previous section.

A similar analysis can be performed in the Ramond sector of the (--1)/(--1)
strings. Here we have sixteen fermionic moduli which we can group into four sets
$M^{\dot\alpha a}$, 
$M^{\alpha\dot a}$, $N^{\alpha a}$ and $N^{\dot\alpha\dot a}$
with $\alpha$, $\dot\alpha$, $a$ and $\dot a$ 
labelling the spinor representations of the four SU(2)'s in (\ref{lorentz}). We have denoted
by $M$'s and $N$'s the components with positive and negative SO(2) chiralities, respectively,
that correspond to eigenvalues $+1$ and $-1$ under $\Omega$ so that the $M$'s, with 
ADHM dimension of (length)$^{+{1}/{2}}$, are in the symmetric
representation 
of $\mathrm{SO}(k)$, while the $N$'s, with dimension of (length)$^{-{3}/{2}}$,
are in the anti-symmetric representation 
of $\mathrm{SO}(k)$. 

\paragraph{(--1)/3 sector} The CP factors of the open strings connecting 
the D-instantons with the D3 brane and its orientifold image are $(k\times 2)$ matrices
transforming in the fundamental representation 
of both $\mathrm{SO}(k)$ and $\mathrm{Sp}(1)$.
The open strings with opposite orientation,
stretching between the D3's and the D(--1)'s, carry CP
factors that are $(2\times k)$ matrices but they should not be counted as independent
because of the identification enforced by the orientifold projection.

The first four directions  of the (--1)/3 strings  have mixed Dirichlet-Neumann
boundary conditions. In the NS sector we therefore find two bosonic moduli, with
ADHM dimension of (length)$^{1}$, in the spinor representation
of $\mathrm{SU}(2)_+$ which we denote by $w_\alpha$. However, if we want to
exhibit also the $\mathrm{SO}(k)$ and $\mathrm{Sp}(1)$ indices, we should write
$w_{\alpha R U}$ (with $R=1,\ldots, k$ and $U=1,2$). For the opposite string
orientation the bosonic
excitations are $\bar w^{\alpha U R}$, but the following orientifold 
identification holds \cite{Gava:1999ky}:
\begin{equation}
 \bar w^{\alpha U R} = \epsilon^{\alpha\beta}\,\epsilon^{UV}\,\delta^{RS}\,
w_{\beta S V}~.
\label{identw}
\end{equation}
In the R sector we find four fermionic moduli, with dimension of
(length)$^{1/2}$, which we denote as $\mu_a$ and $\mu_{\dot a}$ since they are
space-time
scalars but spinors of the internal Lorentz group. The fermionic
moduli $\bar\mu^a$ and $\bar\mu^{\dot a}$ 
arising from open strings with opposite orientation are subject to orientifold identifications 
similar to that displayed in (\ref{identw}).

\paragraph{(--1)/7 sector} In this sector the open string excitations carry
CP factors that are $(k\times 8)$ matrices transforming in the fundamental
representation 
of both $\mathrm{SO}(k)$ and $\mathrm{SO}(8)$. Again the orientifold projection
enforces identifications
with the excitations of open strings with opposite orientations, which
therefore should not be counted 
as independent. Since the (--1)/7 strings have eight directions with mixed Dirichlet-Neumann boundary
conditions, the physical states are only in the R sector where we find one fermionic modulus
$\mu'$ with ADHM dimension of (length)$^{+1/2}$. 
To exhibit also the $\mathrm{SO}(k)$ and $\mathrm{SO}(8)$ indices we should write
$\mu'_{RI}$ (with $R=1,\ldots, k$ and $I=1,\ldots,8$). The physical excitations of the 7/(--1) strings,
whose CP factors are $(8\times k)$ matrices, correspond to the modulus ${\bar \mu}'$ which is subject 
to the following orientifold identification
\begin{equation}
 {\bar \mu}'^{RI} = \delta^{RS}\,\delta^{IJ}\,\mu'_{SJ}~.
\label{identmu'}
\end{equation}
The absence of physical bosonic moduli in this sector is typical of
exotic instanton configurations.

\subsection{BRST structure of moduli space}
The moduli space we have described above admits a BRST structure that will play a crucial
r\^ole for the localization of the integral over the D-instantons. 

The BRST charge $Q$ can be obtained by choosing any component of the supersymmetry charges preserved
on the brane system. The supersymmetry charges are invariant under $\mathrm{SO}(k)$, $\mathrm{Sp}(1)$ and 
$\mathrm{SO}(8)$, but transform as a spinor of the 
$\mathrm{SO}(4)\times \widehat{\mathrm{SO}}(4)$ subgroup 
of the Lorentz group (\ref{lorentz}), 
so that the choice of $Q$ breaks it to the $\mathrm{SU}(2)^3$ 
subgroup which preserves this spinor. In our case we take
\begin{equation}
 \mathrm{SU}(2)_1\times\mathrm{SU}(2)_2\times\mathrm{SU}(2)_3~\equiv~
\mathrm{SU}(2)_-\times\widehat{\mathrm{SU}}(2)_-\times\diag\Big[\mathrm{SU}(2)_+
\times\widehat{\mathrm{SU}}(2)_+\Big]~,
\label{su2s}
\end{equation}
which corresponds to identify the spinor indices $\alpha$ and $a$ of the first 
and third SU(2)'s in (\ref{lorentz}). After this identification is made, the
fermionic moduli $M^{\dot \alpha a}$ and $M^{\alpha\dot a}$ can be renamed as 
$M^{\mu=\dot\alpha \alpha}$ and $M^{m=a\dot a}$, and paired with
$a^\mu$ and $a^m$ into BRST doublets such that $
Qa^\mu= M^\mu$ and $Qa^m=M^m$.
Similarly, in the fermionic sector of the (--1)/(--1) strings
the singlet component $N=\epsilon_{\alpha a}N^{\alpha a}$ has the right transformation properties 
to qualify as BRST partner of $\bar\chi$, {\it i.e.} $Q\bar\chi=N$. Likewise, in the (--1)/3 sector 
the fermionic moduli $\mu_{\alpha=a}$ can be paired with the bosons $w_\alpha$ and form another
BRST doublet such that $Qw_\alpha=\mu_\alpha$.

The remaining fields $N^c = (\tau^c)_{\alpha a}\,N^{\alpha a}$ ($\tau^c$ being the three 
Pauli matrices),
$N^{\dot\alpha\dot a}$, $\mu_{\dot a}$ and $\mu'$ are unpaired, and should be
supplemented with auxiliary fields having identical transformation properties. Denoting
such fields $D^c$, $D^{\dot\alpha\dot a}$, $h_{\dot a}$ and $h'$, respectively, we therefore have
$QN^c = D^c$, $QN^{\dot\alpha\dot a}=D^{\dot\alpha\dot a}$ and so on.

In the usual instanton theory the auxiliary fields collect the D- and F-terms 
of the gauge theory on the D(--1)'s, and the corresponding D- and F-flatness 
conditions are the ADHM constraints on the instanton moduli space 
(see for example \cite{Billo:2002hm,Bruzzo:2002xf,Bruzzo:2003rw} for details).
In our case we have an extension of this construction to a sort of generalized ``exotic'' 
instanton moduli space. More precisely, the seven auxiliary moduli 
$D^c$ and $D^{\dot\alpha\dot a}$, of scaling dimension (length)$^{-2}$
linearize the quartic interactions among the $a^\mu$'s and the $a^m$'s, and in the
explicit string realization correspond to vertex operators that are bi-linear in the 
fermionic string coordinates \cite{Billo:2002hm,Billo:2009di}.
In particular, the triplet $D^c$ disentangles the quartic interactions of the 
$a^\mu$'s and the $a^m$'s among themselves, while the quartet $D^{\dot\alpha\dot a}$
decouples the quartic interactions between the $a^\mu$'s and the $a^m$'s.
Finally, the dimensionless auxiliary moduli $h_{\dot a}$ disentangle the
quartic interactions between $a^m$ and $w_\alpha$, while $h'$ completes the
BRST multiplet in the (--1)/7 sector. In the end only one modulus, namely $\chi$,
remains unpaired and therefore $Q\chi=0$. All this is summarized in Tab. 1.
\begin{table}[ht]
\begin{center}
\begin{tabular}{|c|c|c|c|}
\hline
\phantom{\vdots}sector
& \begin{small} $(\Psi_0,\Psi_1)$ \end{small}
&\begin{small} $\mathrm{SO}(k) \times \mathrm{Sp}(1) \times \mathrm{SO}(8)$ \end{small} 
&\begin{small} $\mathrm{SU}(2)^3$ \end{small}
\\
\hline\hline
$\phantom{\Big|}$D(--1)/D(--1) & $(a^\mu,M^\mu)$ & $\bigl(\Ysymm,\bone,\bone\bigr)$ &
$\repst 212$ \\
$\phantom{\Big|}$\null & $(a^m,M^m)$ & $\bigl(\Ysymm, \bone, \bone\bigr) $
& $\repst 122$ \\
$\phantom{\Big|}$\null & $(N^{\dot\alpha\dot a},D^{\dot\alpha\dot a})$ 
& $\bigl(\Yasymm, \bone, \bone\bigr)$ & $\repst 221$ \\
$\phantom{\Big|}$\null & $(N^{c},D^{c})$ & $\bigl(\Yasymm, \bone, \bone\bigr)$ 
& $\repst 113$ \\
$\phantom{\Big|}$\null & $(\bar\chi,\eta)$ & $\bigl(\Yasymm, \bone, \bone\bigr)$ 
& $\repst 111$ \\
$\phantom{\Big|}$\null & $ \chi$ & $\bigl(\Yasymm, \bone, \bone\bigr)$ 
& $\repst 111$ \\
\hline
$\phantom{\Big|}$D(--1)/D3 & $(w_\alpha,\mu_\alpha)$ & $\bigl(\Yfund,\Yfund ,\bone \bigr) $
& $\repst 112$ \\
$\phantom{\Big|}$\null & $(\mu_{\dot a},h_{\dot a})$ & $\bigl(\Yfund,\Yfund ,\bone\bigr)$ &
$\repst 121$ \\
\hline
$\phantom{\Big|}$D(--1)/D7 & $(\mu',h')$
& $\bigl(\Yfund, \bone, \Yfund\bigr)$
& $\repst 111$ \\
\hline
\end{tabular}
\end{center}
\label{tab:mod_spec}
\caption{Spectrum of moduli for the D7/D3/D(--1) system in type I', arranged in
BRST doublets $(\Psi_0,\Psi_1)$ such that $Q\Psi_0=\Psi_1$. The
last column displays the representations of the various moduli under the three SU(2)'s 
defined in (\ref{su2s}).}
\end{table}

Since the scaling dimension of the BRST charge is (length)$^{-1/2}$, the dimensions of the
components $(\Psi_0,\Psi_1)$ of any BRST doublet are of the form (length)$^\Delta$ and (length)$^{\Delta-1/2}$. Thus, recalling
that a fermionic variable and its differential have opposite dimensions, 
the measure on the instanton moduli space 
\begin{equation}
 d\mathcal{M}_{k} \equiv \,d\chi \prod_{(\Psi_0,\Psi_1)} d\Psi_0\,d\Psi_1
\label{measure}
\end{equation}
has the total dimension
\begin{equation}
 \mbox{(length)}^{-\frac{1}{2}k(k-1)+\frac{1}{2}n_b-\frac{1}{2}n_f}~.
\label{dimmeas}
\end{equation}
Here, the first term in the exponent accounts for the unpaired modulus $\chi$ in the
anti-symmetric representation of $\mathrm{SO}(k)$, while $n_b$ ($n_f$) denotes the number
of BRST multiplets whose lowest components $\Psi_0$ are bosonic (fermionic). From
Tab.~1 it is not difficult to verify that
\begin{equation}
 n_b= \frac{9}{2}k^2+\frac{15}{2}k~~\mbox{and}~~n_f=\frac{7}{2}k^2+\frac{17}{2}k~.
\end{equation}
Hence, \eq{dimmeas} implies that the full measure $d\mathcal{M}_{k}$ 
is dimensionless, in agreement with
the conformality of the theories on both the D3's and the D7's.

For all the moduli we have listed above, 
including the auxiliary ones, it is possible to write vertex operators of conformal 
dimension one and use them to obtain the moduli action 
$\mathcal{S}(\mathcal{M}_{k})$ by computing disk amplitudes along the lines discussed in 
\cite{Green:2000ke,Billo:2002hm,Billo:2006jm,Billo:2009di}. The result is the moduli
action derived in \cite{Gava:1999ky} (see in particular Eq. (3.3) of that
paper).
A first generalization
that will be necessary in the following is the introduction of the interaction
terms between the moduli and the Sp(1) adjoint scalar $\phi$ of the 3/3 sector
or the SO(8) adjoint scalar $m$ of the 7/7 sector.
As explained for example in \cite{Billo:2006jm},
such interaction terms can be derived from mixed disk amplitudes
with a portion of the boundary on the D3's or the D7's where a scalar field
$\phi$ or $m$ can be emitted. The result of these computations is the action
$\mathcal{S}(\mathcal{M}_{k};\phi,m)$. It turns out that this can be
obtained from $\mathcal{S}(\mathcal{M}_{k})$ with a
simple effective rule corresponding to a shift of the $\chi$ moduli.
More precisely, we have to make the following replacements%
\footnote{As noted in footnote \ref{foot:m}, in our conventions it is the field
$\ii m$ which
is in the adjoint of SO(8).}
\begin{equation}
\begin{aligned}
 \epsilon^{UV}\,\chi^{RS} &~\to~\epsilon^{UV}\,\chi^{RS} +\phi^{UV}\,\delta^{RS}~,\\
 \delta^{IJ}\,\chi^{RS} &~\to~\delta^{IJ}\,\chi^{RS} +\ii\, m^{IJ}\,\delta^{RS}~.\\
\end{aligned}
\label{shifts}
\end{equation}
Furthermore, one can prove that this action is BRST exact, namely
\begin{equation}
\mathcal{S}(\mathcal{M}_{k};\phi,m) = Q\,\Xi
\label{exactness}
\end{equation}
for a suitably defined fermion $\Xi$ and that
$Q$ is nilpotent up to infinitesimal transformations of $\mathrm{SO}(k)$, $\mathrm{Sp}(1)$
and $\mathrm{SO}(8)$ parameterized respectively by $\chi$, $\phi$ and $\ii\,m$. This means that
on any modulus we have
\begin{equation}
 Q^2 \,\bullet = \big[T_{\mathrm{SO}(k)}(\chi) +
T_{\mathrm{Sp}(1)}(\phi) + T_{\mathrm{SO}(8)}(\ii\,m)\big]\,\bullet~,
\label{Q2}
\end{equation}
where the $T$'s act in the specific $\mathrm{SO}(k)$, $\mathrm{Sp}(1)$
and $\mathrm{SO}(8)$ representations given in Tab.~1. 

For our later purposes it
is enough to consider the Cartan directions of the various groups. We label the Cartan
components of the $\mathrm{SO}(k)$ parameters of $Q$ by $\vec \chi=\{\chi_r\}$ with $r=1,\ldots,
\mathrm{rank}\,\mathrm{SO}(k)$, those of Sp(1) by $\vec\phi=\{a\}$, and those
of SO(8) by $\vec m=\{{m_i}/{\sqrt{2}}\}$ 
with $i=1,\ldots,4$. The latter two choices are in agreement with the choices and
normalizations used in the previous section, see in particular Eq.s~(\ref{defa})
and (\ref{psitom}). Using these ingredients, one can easily see that $Q^2$ in
(\ref{Q2}) corresponds to infinitesimal Cartan actions which can be diagonalized
in any representation by going to the basis provided by the weights.
Indeed, denoting respectively by $\vec\pi$, $\vec\sigma$ and $\vec\rho$ the weights of the
$\mathrm{SO}(k)$, $\mathrm{Sp}(1)$ and $\mathrm{SO}(8)$ representations under which a
given modulus transforms, we can rewrite (\ref{Q2}) as
\begin{equation}
Q^2 \,\bullet= \big[\,\vec\chi\cdot\vec\pi+\ii\,\vec\phi\cdot\vec\sigma+\ii\,\vec m\cdot\vec\rho
\,\big]\bullet~.
\label{weight}
\end{equation}
To fully localize the integral over moduli space it is necessary to use a BRST
charge that is equivariant with respect to all symmetries of the model. In our case these
include also the residual Lorentz symmetry $\mathrm{SU}(2)^3$ defined in
(\ref{su2s}), besides the $\mathrm{SO}(k)$, $\mathrm{Sp}(1)$
and $\mathrm{SO}(8)$ symmetries considered so far. 

\subsection{Moduli integration via localization and instanton partition function}
\label{subsec:modint}
As explained for example in \cite{Nekrasov:2002qd}--\nocite{Bruzzo:2002xf,Nekrasov:2003rj}\cite{Bruzzo:2003rw}, 
the fully equivariant cohomology can be obtained by deforming the moduli action with parameters
associated to ``rotations'' along the Cartan directions of the residual Lorentz group. For our model,
these rotations can be parameterized by $\vec\epsilon=\{\epsilon_A\}$ with $A=1,\ldots,4$, 
subject to the constraint
\begin{equation}
 \epsilon_1+\epsilon_2+\epsilon_3+\epsilon_4=0~.
\label{constr}
\end{equation}
Although only three out of the $\epsilon$'s are independent variables, 
in all calculations it is convenient to use all four of them and impose the relation (\ref{constr})
only at the very end. From the string point of view the $\epsilon$-deformation can be
obtained by switching on a RR background on the D7 brane world-volume corresponding to
graviphoton field strengths of the type
\begin{equation}
\mathcal F_{\mu\nu} \sim \begin{pmatrix} 
      0& +\epsilon_1 &0&0  \cr
     -\epsilon_1&0&0&0 \cr
      0&0&0& +\epsilon_2\cr
      0&0&-\epsilon_2&0 
     \end{pmatrix} 
~~~~\mbox{and}~~~~\mathcal F_{mn} \sim \begin{pmatrix} 
      0& +\epsilon_3 &0&0  \cr
     -\epsilon_3&0&0&0 \cr
      0&0&0& +\epsilon_4\cr
      0&0&-\epsilon_4&0 
     \end{pmatrix}~.
\label{graviphoton}
\end{equation}
This RR background induces new $\epsilon$-dependent terms in the moduli action (\ref{exactness})
which then gets replaced by a deformed action $\mathcal{S}(\mathcal{M}_{k};\phi,m,\epsilon)$. 
The $\epsilon$-dependent terms can be explicitly derived by computing
mixed open/closed string amplitudes on disks with insertions of the moduli vertex operators
on the boundary and of the vertex operators representing the graviphoton field
strengths $\mathcal F$ in the interior, as shown in \cite{Billo:2006jm,Billo:2009di,Ito:2010vx} 
for similar systems.  The net result of these computations amounts to replace
the old BRST charge $Q$ with a deformed charge $\widetilde Q$ satisfying the following relation
\begin{equation}
{\widetilde Q}^2 \,\bullet= 
\big[\,\vec\chi\cdot\vec\pi+\ii\,\vec\phi\cdot\vec\sigma+\ii\,\vec m\cdot\vec\rho + 
\vec\epsilon \cdot \vec \gamma\,\big]\bullet~,
\label{weight1}
\end{equation}
which is a direct generalization of (\ref{weight}). Here we have taken 
the deformation parameters $\epsilon_A$ to be of dimensions
(length)$^{-1}$ and have denoted by $\vec \gamma$ the weights of the
representation of the residual Lorentz group (\ref{su2s}) under which a given modulus
transforms. Using the rules explained in \cite{Billo':2010bd}, one can show that
\begin{equation}
 \vec\epsilon\cdot \vec \gamma = (\epsilon_1-\epsilon_2)\gamma_1+
(\epsilon_3-\epsilon_4)\gamma_2+(\epsilon_1+\epsilon_2)\gamma_3~,
\label{qdotepsilon}
\end{equation}
where $\gamma_\ell$ are the weights of the three SU(2)'s, namely 
$\gamma_\ell=0$ for a modulus in the $\mathbf{1}$, $\gamma_\ell=\pm\frac{1}{2}$
for a modulus in the $\mathbf{2}$ and so on. The explicit values of 
$\vec\epsilon\cdot \vec \gamma$ for all BRST doublets of moduli are collected in
Tab.~2.
\begin{table}[ht]
\begin{center}
\begin{tabular}{|c|c|c|}
\hline
\phantom{\vdots}sector
& \begin{small} $(\Psi_0,\Psi_1)$ \end{small}
&\begin{small} $\vec\epsilon\cdot \vec \gamma$ \end{small}
\\
\hline\hline
$\phantom{\Big|}$D(--1)/D(--1) & $(a^\mu,M^\mu)$ &
$\pm\epsilon_1$, $\pm\epsilon_2$ \\
$\phantom{\Big|}$\null & $(a^m,M^m)$ 
& $\pm\epsilon_3$, $\pm\epsilon_4$\\
$\phantom{\Big|}$\null & $(N^{\dot\alpha\dot a},D^{\dot\alpha\dot a})$ 
& $\pm(\epsilon_2 +\epsilon_3)$, $\pm(\epsilon_1+\epsilon_3)$\\
$\phantom{\Big|}$\null & $(N^{c},D^{c})$ 
& $0$, $\pm(\epsilon_1 +\epsilon_2)$ \\
$\phantom{\Big|}$\null & $(\bar\chi,\eta)$ & $0$ \\
\hline
$\phantom{\Big|}$D(--1)/D3 & $(w_\alpha,\mu_\alpha)$ 
& $\pm\frac{1}{2}(\epsilon_1+\epsilon_2)$ \\
$\phantom{\Big|}$\null & $(\mu_{\dot a},h_{\dot a})$ &
$\pm\frac{1}{2}(\epsilon_3-\epsilon_4)$ \\
\hline
$\phantom{\Big|}$D(--1)/D7 & $(\mu',h')$
& $0$ \\
\hline
\end{tabular}
\end{center}
\label{tab:epsilon_weight}
\caption{The last column displays the values of $\vec\epsilon\cdot\vec \gamma$ for the various
BRST doublets.}
\end{table}

The $\epsilon$-deformed moduli action
is still BRST exact (with respect to the new BRST charge $\widetilde Q$), namely
\begin{equation}
 \mathcal{S}(\mathcal{M}_{k};\phi,m,\epsilon) = \widetilde Q\,\widetilde \Xi
\label{sqepsilon}
\end{equation}
for a suitable fermion $\widetilde \Xi$. This deformed BRST structure allows to
perform rescalings of various moduli and show that the $k$-instanton partition
function
\begin{equation}
 \mathcal{Z}_k~
\equiv~\mathcal{N}_k\int d\mathcal{M}_{k}~\ee^{-\mathcal{S}(\mathcal{M}_{k};\phi,m,\epsilon)}
\label{zk}
\end{equation}
(with $\mathcal N_k$ a normalization factor)
can be computed exactly in the semiclassical approximation.
Indeed, the complete localization of the integral over moduli space
around isolated fixed points implies
that $\mathcal{Z}_k$ is given by  the (super)-determinant of ${\widetilde Q}^2$ 
evaluated at the fixed points of $\widetilde Q$ 
\cite{Nekrasov:2002qd}--\nocite{Bruzzo:2002xf,Nekrasov:2003rj}\cite{Bruzzo:2003rw}.
As we already mentioned, the moduli $\chi$ and $\bar\chi$ appear very
asymmetrically in the BRST formalism: $\chi$ parametrizes the $\mathrm{SO}(k)$ rotations,
while $\bar\chi$ falls into one of the BRST doublets. Moreover, 
the contribution of the $(\bar\chi,\eta)$ multiplet to the super-determinant
cancels against an identical contribution coming from the neutral component in
$(N^c,D^c)$ with identical transformation properties and opposite statistics.
Taking this into account, the super-determinant of ${\widetilde Q}^2$ takes a simple product 
form in terms of the $\widetilde Q^2$-eigenvalues given in (\ref{weight1})
and the $k$-instanton partition function (\ref{zk}) can be expressed
by the localization formula
\begin{equation}
 {\mathcal Z}_k  = \cN_k \int ~\prod_{r=1}^{\mathrm{rank}\,\mathrm{SO}(k)} 
\Big(\frac{d\chi_r}{2\pi\ii}\Big) ~\Delta(\chi)~ 
 \frac{P_{\Yasymm}(\chi)}{P_{\Ysymm}(\chi)} ~P_{\Yfund}(\chi)~.
\label{zkloc}
\end{equation}
Here $\Delta(\chi)$ is the Vandermonde determinant representing the Jacobian
factor for the diagonalization of the integration variables $\chi$.
The factor $P_{\Yasymm}(\chi)$ arises from the integration over the 
BRST doublets containing $N^c$ and $N^{\dot\alpha\dot a}$ which transform in
the $\Yasymm$ representation of $\mathrm{SO}(k)$. It is given by
\begin{equation}
 \label{padj}
  P_{\Yasymm}(\chi) = (-s_1 s_2 s_3)^{r_{\Yasymm}}~
  \prod_{\ell=1}^3 \Bigg\{\prod_{\vec\pi\in{\Yasymm}^+} \Big[\,(\vec\chi\cdot\vec\pi)^2-
 s_\ell^2\,\Big]\Bigg\}~,
\end{equation}
where we have introduced the following combinations
\begin{equation}
 \label{sdef}
  s_1 = \epsilon_2 + \epsilon_3~,~~~
  s_2 = \epsilon_1 + \epsilon_3~,~~~
  s_3 = \epsilon_1 + \epsilon_2~,
\end{equation}
and denoted by $r_{\Yasymm}$ the number of null weights in the anti-symmetric representation
of SO$(k)$ (namely, the rank of the group) and by ${\Yasymm}^+$ the set of its positive weights.

The factor $P_{\Ysymm}(\chi)$ in the denominator of (\ref{zkloc})
arises from the integration over the BRST doublets containing $a^\mu$ and $a^m$, and 
is given by
\begin{equation}
 \label{psym}
  P_{\Ysymm}(\chi) = (\epsilon_1\epsilon_2\epsilon_3\epsilon_4)^{r_{\Ysymm}}~ 
  \prod_{A=1}^4 \Bigg\{\prod_{\vec\pi\in{\Ysymm}^+} \Big[\,(\vec\chi\cdot\vec\pi)^2-
  \epsilon_A^2\,\Big]\Bigg\}~,
\end{equation}
where again $r_{\Ysymm}$ is the number of null weights in the symmetric 
representation of SO$(k)$ and ${\Ysymm}^+$ is the set of its positive weights. 

Finally, the last factor $P_{\Yfund}(\chi)$ in (\ref{zkloc})
collects the contributions from the integration over the BRST doublets 
containing the moduli with only one end-point on the D-instantons, namely 
the $\mu'$'s from the (--1)/7 sector, and the
$w_\alpha$'s and $\mu_{\dot a}$'s from the (--1)/3 sector. It reads
\begin{equation}
 \label{Pfun}
 \begin{aligned}
  P_{\Yfund}(\chi) & = \left(\frac{\Pf m}{4}~ \frac{a^2 + (\frac{\epsilon_3 -
  \epsilon_4}{2})^2}{a^2 + (\frac{\epsilon_1+
  \epsilon_2}{2})^2}\right)^{r_{\Yfund}}
  \,\prod_{\vec\pi\in{\Yfund}^+}\Bigg\{ \prod_{i=1}^4\Big[\,(\vec\chi\cdot\vec\pi)^2 +
  \frac{m_i^2}{2}\big)\,\Big]\\
  & ~~~~~\times 
  \frac{\Big[\,(\vec\chi\cdot\vec\pi)^2 - \bigl(\ii\,a +
  \frac{\epsilon_3 -\epsilon_4}{2} \bigr)^2\,\Big]
  \,\Big[\,(\vec\chi\cdot\vec\pi)^2 - \bigl(\ii\,a -
  \frac{\epsilon_3 - \epsilon_4}{2} \bigr)^2\,\Big]}
{\Big[\,(\vec\chi\cdot\vec\pi)^2 - \bigl(\ii\,a +
  \frac{\epsilon_1 +\epsilon_2}{2} \bigr)^2\,\Big]
  \,\Big[\,(\vec\chi\cdot\vec\pi)^2 - \bigl(\ii\,a -
  \frac{\epsilon_1 + \epsilon_2}{2} \bigr)^2\,\Big]}\Bigg\}~,
 \end{aligned}
\end{equation}
where $r_{\Yfund}$ is the number of null weights in the fundamental
representation of SO$(k)$ and ${\Yfund}^+$ the set of its positive weights.

More explicit expressions for $\Delta(\chi)$, $P_{\Yasymm}(\chi)$, $P_{\Ysymm}(\chi)$ 
and $P_{\Yfund}(\chi)$ are given in Appendix~\ref{app:details}. Here we just remark 
that the integrand in (\ref{zkloc}) is a rational function and that the integrals over
$\chi_r$ have to be computed according to the prescription of \cite{Moore:1998et}
as contour integrals after giving the deformation parameters $\epsilon_A$'s a positive
imaginary part such that
\begin{equation}
 \im \epsilon_1 > \im \epsilon_2>\cdots>\im \epsilon_4>\im \frac{\epsilon_1}{2}>\cdots
>\im \frac{\epsilon_4}{2}~.
\label{prescr}
\end{equation}
In this way the calculation of the D-instanton partition function $\mathcal{Z}_k$ is reduced
to the sum of the residues of a rational function with simple poles.

\subsection{Non-perturbative prepotentials}
\label{subsec:prepot}

In order to derive the non-perturbative contributions to the world-volume effective actions of the
space-filling branes we have to sum over all instanton numbers and consider the
``grand-canonical'' instanton partition function
\begin{equation}
 Z = \sum_{k=0}^\infty  \mathcal{Z}_k\,q^k~,
\label{partfun}
\end{equation}
where we have conventionally set $\mathcal{Z}_0=1$. Then, we have to switch off the $\epsilon$-deformations. However, in doing this we have to pay attention to a couple of
points. In fact, as is clear from (\ref{zk}), the integrals appearing in $\mathcal{Z}_k$ run
over {\emph{all}} moduli, including also the ``center of mass'' super-coordinates $X$ and $\Theta$.
In presence of the $\epsilon$-deformations it is rather easy to see that the integration
over this eight-dimensional super-space yields a volume factor growing as $1/(\epsilon_1\epsilon_2\epsilon_3
\epsilon_4)$ in the limit of small $\epsilon_A$'s. Therefore, to obtain the integral
over the centered moduli and hence the contributions to the brane effective actions, 
this volume factor has to be removed before turning off the $\epsilon$-deformations.
In addition, we have to take into account the fact that the $k$-th order in the $q$-expansion 
receives contributions not only from genuine $k$-instanton configurations 
but also from disconnected ones, corresponding to copies of instantons of 
lower numbers $k_\ell$ such that $\sum_\ell k_\ell = k$. To isolate the connected
components we have to take the logarithm of $Z$. 

The singularity structure of $\log Z$ with respect to the $\epsilon$-deformations
turns out to be the following:
\begin{equation}
 \label{singlogZ}
 \log Z = \frac{1}{\epsilon_1\epsilon_2\epsilon_3\epsilon_4}
 {\widehat\cF}^{(8)}_{\mathrm{n.p.}} +
 \frac{1}{\epsilon_1\epsilon_2} {\widehat\cF}^{(4)}_{\mathrm{n.p.}}~.
\end{equation}
Here, ${\widehat\cF}^{(8)}_{\mathrm{n.p.}}$ and ${\widehat\cF}^{(4)}_{\mathrm{n.p.}}$
are expressions which are regular when the deformations are switched off;
the subscript refers to the non-perturbative nature of these quantities, while the
superscripts signal their eight-dimensional, respectively four-dimensional,
nature. Since the $\epsilon_A$'s have dimension of (length)$^{-1}$,
${\widehat\cF}^{(8)}_{\mathrm{n.p.}}$ must have mass-dimension four, while 
${\widehat\cF}^{(4)}_{\mathrm{n.p.}}$ must have mass-dimension two. Indeed, ${\widehat\cF}^{(8)}_{\mathrm{n.p.}}$ is a quartic expression in $m_i$ and $\epsilon_A$,
without any dependence on $a$. Therefore we can conclude that ${\widehat\cF}^{(8)}_{\mathrm{n.p.}}$
represents the non-perturbative quartic prepotential for the D7 brane world-volume theory.
On the other hand, ${\widehat\cF}^{(4)}_{\mathrm{n.p.}}$ is quadratic 
in $\epsilon_A$, $a$ and $m_i$; since
$1/(\epsilon_1\epsilon_2)$ represents the regulated (super)volume in the first four
directions, we can conclude that ${\widehat\cF}^{(4)}_{\mathrm{n.p.}}$
is the non-perturbative quadratic prepotential for the D3 brane world-volume
theory%
\footnote{Note that terms in $\widehat \cF^{(8)}_{\mathrm{n.p.}}$ of the form
$\epsilon_3\epsilon_4\,f$ could in principle
be interpreted also as terms in $\widehat \cF^{(4)}_{\mathrm{n.p.}}$
proportional to $f$. We fix this kind of ambiguity by requiring that all terms assigned to
$\widehat \cF^{(8)}_{\mathrm{n.p.}}$ have the structure and symmetry properties that are
appropriate for a correct eight-dimensional interpretation as a prepotential.}. 

The prepotentials ${\widehat\cF}^{(8)}_{\mathrm{n.p.}}$ and 
${\widehat\cF}^{(4)}_{\mathrm{n.p.}}$ contain ``gravitational'' corrections
encoded in the deformation parameters $\epsilon_A$ which, as we
have described in Section \ref{subsec:modint}, arise from the RR closed string
sector. In this paper we are not interested in this dependence, and we limit 
ourselves to the prepotentials
\begin{equation}
 \label{limprep}
  \cF^{(8)}_{\mathrm{n.p.}} = \lim_{\epsilon_A\to 0}
  {\widehat\cF}^{(8)}_{\mathrm{n.p.}}~,~~~
  \cF^{(4)}_{\mathrm{n.p.}} = \lim_{\epsilon_A\to 0}
  {\widehat\cF}^{(4)}_{\mathrm{n.p.}}
 \end{equation}
which depend on the gauge theory parameters only.

It should not be a surprise that the D-instanton partition function $Z$ in presence
of D7 and D3 branes contains both an eight-dimensional and a four-dimensional information. 
Indeed, depending on the point of view, the D(--1)'s can be considered alternatively 
either as (exotic) instantons for the eight-dimensional theory on the D7's or as
(ordinary) instantons for the four-dimensional theory on the D3's. In the following sections
we will consider in detail the non-perturbative prepotentials (\ref{limprep}),
give their explicit expressions for the first instanton numbers (up to $k=5$),
analyze their properties and discuss their relations with the 
F-theory perspective and the SW curve we discussed in the previous
section.

\section{Eight-dimensional effective prepotential and chiral ring}
\label{sec:8d}
The non-perturbative prepotential for the eight-dimensional gauge theory on the D7 branes 
can be computed using formulae (\ref{singlogZ}) and (\ref{limprep}). Up to
instanton order $k=5$, we find
\begin{equation}
 \label{prep8d}
 \begin{aligned}
\phantom{\Bigg[}  \cF_{\rm n.p.}^{(8)} & = -2\,\Pf m 
   \Big(q + \frac 43 q^3 + \frac 65 q^5 + \ldots\Big)\\
  &~~~ \,- \frac 12 \sum_{i<j} m_{i}^2 m_{j}^2\, q^2 - \frac 18
  \Big(\sum_i m_{i}^4 + 4 \sum_{i<j} m_{i}^2 m_{j}^2\Big)\, q^4 + \ldots~.
 \end{aligned}
\end{equation}
This prepotential is independent on the vacuum expectation value $a$ of the D3 adjoint
multiplet, and coincides exactly%
\footnote{Notice that our conventions are changed with respect to \cite{Billo:2009di}
in that we now have $q=\exp(\pi\ii\tau_0)$ to agree with the conventions
of \cite{Seiberg:1994rs}, while there we wrote $q=\exp(2\pi\ii\tau)$. Moreover, we
choose the overall normalization of the instanton partition function 
in a way which corresponds to sending $q \to -q$, {\it i.e.} to changing the sign in 
front of the Pfaffian terms. Finally, the field $m$ we use here corresponds to the field $\phi$ 
of \cite{Billo:2009di}; however, because of our conventions in \eq{psitom}, the
eigenvalues $\phi_i$  of the latter correspond to $-\ii m_i/\sqrt{2}$.} with the
result found in \cite{Billo:2009di} for the D7/D(--1) system in type I$^\prime$.
Using the quadratic and quartic SO(8) invariants defined in 
Appendix~\ref{app:flavinv},
we can rewrite the prepotential (\ref{prep8d}) as follows
\begin{equation}
 \label{prep8d2}
 \begin{aligned}
  \cF_{\mathrm{n.p.}}^{(8)} & = 2 \,T_1 \left(q - \frac 32 q^2 + \frac 43 q^3 - \frac 34 q^4
+ \frac 65 q^5 + \ldots\right)
\\  &~~~~+
4 \,T_2 \left(q + \frac 43 q^3 + \frac 65 q^5 + \ldots\right)
-R^2 \left( \frac 12 q^2 + \frac 34 q^4 + \ldots\right) 
 ~.
 \end{aligned}
\end{equation}
This non-perturbative part has to be added to the tree-level prepotential
\begin{equation}
 \cF_{\mathrm{tree}}^{(8)} = \frac{\pi\ii\tau_0}{4!}\,\Tr m^4 =
-\frac{1}{4}\,T_1\,\ln q + \frac{1}{24}\,R^2\,\ln q~,
\label{tree}
\end{equation}
and to the perturbative 1-loop contribution which, up
to terms that depend on the complex structure $U$ of the torus transverse to the D7 branes
and that we do not write for brevity, is
given by \cite{Billo':2009gc}
\begin{equation}
 \cF_{\mathrm{1-loop}}^{(8)} = \frac{1}{32}\,\ln \big( \frac{\im \tau_0}{2}\big)\, \big(\Tr m^2\big)^2
= \frac{1}{8}\,\ln \big( \frac{\im \tau_0}{2}\big)\, R^2~.
\label{1loop}
\end{equation}

As shown in \cite{Billo:2009di}, the above results up to order $k=5$ are consistent 
with heterotic/type I$^\prime$ duality \cite{Lerche:1998nx,Gava:1999ky,Gutperle:1999xu,Kiritsis:2000zi}
and may be extended to all orders in $q$. Indeed, the complete prepotential
may be written as
\begin{equation}
\begin{aligned}
 \phantom{\Bigg\{}
\cF^{(8)}  &= \cF_{\mathrm{tree}}^{(8)} + \cF_{\mathrm{1-loop}}^{(8)} + \cF_{\mathrm{n.p.}}^{(8)}\\
&=-\frac{1}{4}\,T_1\,\ln \Big(\frac{\kappa(q)}{16}\Big)
-\frac{1}{4}\,T_2\,\ln \Big(1-\kappa(q)\Big)
+\frac{1}{8}\,R^2 \ln\Big(\frac{\im \tau_0}{2}\,\eta^4(\tau_0)\Big)~,
\end{aligned}
 \label{preptot1}
\end{equation}
where
\begin{equation}
 \kappa(q) = \frac{\vartheta_2^4(\tau_0)}{\vartheta_3^4(\tau_0)}~.
\label{kappa}
\end{equation}
Expanding the $\vartheta$-functions and the Dedekind function in powers of $q$, one can
easily check the agreement with the perturbative and non-perturbative expressions reported
above.
Up to $\tau_0$-independent terms that can be presumably absorbed
by a (finite) renormalization of the 1-loop contribution (\ref{1loop}) which was not considered
in \cite{Billo:2009di}, we can further rewrite the eight-dimensional prepotential as follows
\begin{equation}
\cF^{(8)} = -\frac{1}{4}\,T_1\,\ln \big(\kappa(q)\big)
-\frac{1}{4}\,T_2\,\ln \big(1-\kappa(q)\big)
+\frac{1}{8}\,R^2 \ln\big(\im \tau_0\,\eta^4(\tau_0)\big)~.
 \label{preptot2}
\end{equation}
Notice that the single-trace quartic structure present in $T_1$, whose 
tree-level coupling is $\pi\ii\tau_0$ (see \eq{tree}), has an 
effective coupling given by 
\begin{equation}
 \label{ef8}
 \ln \kappa(q) = \ln q + \ln 16 - 8 q + 12 q^2 - \frac{32}{3} q^3 + 
6q^4 -\frac{48}{5} q^5 + \ldots~.
\end{equation}
This single-trace structure is conformal at the perturbative level
but from (\ref{ef8}) we see that it receives both a finite 1-loop renormalization (the $\ln 16$ term above) and a series of instanton contributions. In Section~\ref{sec:compare} we will find again
this same non-perturbative redefinition (or more precisely its inverse) in the four-dimensional SU(2) SYM theory living on the D3 branes.

{From} the prepotential $\cF^{(8)}$ we obtain the eight-dimensional SO(8) effective action
on the D7 branes by promoting the diagonal vacuum expectation values $m_i$ to the full
chiral superfield $M(X,\Theta)$ given in (\ref{Phi8}) and integrating over the eight-dimensional chiral
superspace, namely
\begin{equation}
 \label{Seff8}
  S^{(8)} = \frac{1}{(2\pi)^4}\int d^8X\, d^8\Theta ~ \cF^{(8)}(M)
  ~+~ \mathrm{c.c.}~.
\end{equation}
Among various other terms, this effective action contains quartic couplings of the type $\Tr\big(t_8\,f^4\big)$ with
$t_8$ being the totally anti-symmetric eight-index tensor appearing in various
string amplitudes \cite{Green:1987mn} and $f$ being the field strength, that under the
heterotic/type I$^\prime$ duality exactly match the heterotic results
\cite{Billo:2009di}.

As we discussed in Section~\ref{subsec:FD3}, 
the mathematical description of the non-perturbative effects in the eight-dimensional type I$^\prime$ 
theory should be identical to the non-perturbative description of the
four-dimensional $\cN=2$ SYM theory with gauge group SU(2) and $N_f=4$ (massive) flavors.
For such a theory, which possesses a global SO(8) flavor symmetry, it was argued in \cite{Seiberg:1994rs} that the full symmetry of the effective action 
is $\mathrm{SL}(2,\mathbb{Z})$ whose generators act as standard modular transformations 
on the tree-level gauge coupling accompanied by triality transformations 
on the flavor SO(8) representations. Therefore, following the arguments of \cite{Sen:1996vd},
we expect that such $\mathrm{SL}(2,\mathbb{Z})$ be a symmetry of the eight-dimensional effective 
action (\ref{Seff8}), in which now SO(8) is the gauge group. 
We now show that this is indeed the case. 
For simplicity we assume that the background axionic value be vanishing,
{\it i.e.} we take $\re \tau_0=0$. 

Let us first consider the transformation $\widehat{\mathcal S}$ defined as
\begin{equation}
 \widehat{\mathcal S} ~:~~\tau_0\longrightarrow -\frac{1}{\tau_0}~~,~~
T_1\longleftrightarrow T_2~.
\label{s} 
\end{equation}
This is the combination of the $\mathcal S$ modular transformation on the coupling constant $\tau_0$
with the SO(8) triality transformation that exchanges the vector representation (associated
to $T_1$) with one of the spinorial representations (associated to $T_2$) leaving fixed
the other (associated to $T_3=-T_1-T_2$) as well as the quadratic invariant $R$. Using the modular properties of the Jacobi $\vartheta$-functions (see \eq{Smodprop}), it is easy to see that
\begin{equation}
 \kappa(q) \longleftrightarrow {1-\kappa(q)}~,
\label{kappa_s}
\end{equation}
and hence that the first two terms of the prepotential (\ref{preptot2}) are invariant
under $\widehat{\mathcal S}$. Furthermore, for the case at hand we have
\begin{equation}
 \im \tau_0 \,\eta^4(\tau_0) \longrightarrow -\frac{\tau_0}{\bar\tau_0}~\im \tau_0 \,\eta^4(\tau_0)
=\im \tau_0 \,\eta^4(\tau_0)~,
\label{s1}
\end{equation}
so that also the $R^2$ term in (\ref{preptot2}) is invariant under $\widehat{\mathcal S}$.

Now let us consider the transformation $\widehat{\mathcal T}$ defined as
\begin{equation}
 \widehat{\mathcal T} ~:~~\tau_0\longrightarrow \tau_0 +1~~,~~T_2\longleftrightarrow T_3=-T_1-T_2~.
\label{t} 
\end{equation}
This is the combination of the $\mathcal T$ transformation acting on $\tau_0$ with the 
triality transformation that exchanges the two spinor representations of SO(8), leaving fixed the
vector representation (and the quadratic invariant $R$). {From} the modular properties under $\cT$
given in (\ref{Tmodprop}), it is easy to check that
\begin{equation}
\kappa(q) \longrightarrow \ee^{\pi\ii}\,\frac{\kappa(q)}{1-\kappa(q)}~~,~~
1-\kappa(q) \longrightarrow \frac{1}{1-\kappa(q)}~~,~~
\im \tau_0 \,\eta^4(\tau_0) \longrightarrow \ee^{\frac{\pi\ii}{3}}\,\im \tau_0 \,\eta^4(\tau_0)~,
\label{kappa_t}
\end{equation}
and hence that the prepotential (\ref{preptot2}) transforms under $\widehat{\mathcal T}$ as follows 
\begin{equation}
  \cF^{(8)} \longrightarrow
\cF^{(8)}-\frac{\pi\ii}{4}\,T_1+\frac{\pi\ii}{24}\,R^2 ~=~
\cF^{(8)}+\frac{\pi\ii}{4!}\,\Tr m^4~.
\label{prepot_t}
\end{equation}
Even if the prepotential is not invariant, the change produced 
in the effective action is unobservable because
\begin{equation}
 \begin{aligned}
  \delta S^{(8)} &=\frac{1}{(2\pi)^4}\,\int d^8X\,d^8\Theta ~\delta 
\cF^{(8)}(M)
+ ~\mathrm{c.c.}\\
&=\frac{\pi\ii}{4!(2\pi)^4}\,\int d^8X\,d^8\Theta ~\big[\Tr M^4-\Tr {\overline M}^{\,4}\big] 
~=~ 2\pi\ii \,c_{4}~,
 \end{aligned}
\label{change}
\end{equation}
where in the last step $c_4$ is the topological index of order four which is an integer number.
Since the effective action shifts by an integer multiple of $2\pi\ii$, we can conclude%
\footnote{This is the same argument used to show that the classical
tree-level action following from (\ref{tree}) is invariant under $\tau_0~\to~\tau_0+1$.}
that the theory is invariant also under $\widehat{\mathcal T}$. Thus, the 
symmetry group of the quantum SO(8) gauge theory living on the D7 branes is the one generated
by $\widehat{\mathcal S}$ and $\widehat{\mathcal T}$, which, as shown in Appendix~\ref{subsec:trsl}, 
is isomorphic to $\mathrm{SL}(2,\mathbb{Z})$ in agreement with our expectations.

Besides the non-perturbative effective action and its symmetry properties, another useful 
set of information on the eight-dimensional theory is provided by the quantum
vacuum expectation values of the composite operators $\Tr m^{2l}$, forming the 
so-called ``chiral ring'' of the model. These have been computed in \cite{Fucito:2009rs}
via localization techniques with suitable insertions in the instanton partition function
for a generic number of D7 branes. When this number is four, {\it i.e.} for our SO(8)
gauge theory, the results for the first
few values of $l$ are%
\footnote{With respect to \cite{Fucito:2009rs}, our present conventions are such that
their $\Tr \phi^{2l}$ is mapped to $(-1)^l 2^l\, \Tr m^{2l}$ and their $q$ goes into
our $(-q)$.}
\begin{equation}
\begin{aligned}
 \label{chirexp}
 \vev{\Tr m^2} & = \Tr \vev{m}^2\phantom{\Big(}~,\\
 \vev{\Tr m^4} & = \Tr \vev{m}^4 - 48\, 
\Pf m ~q -24\sum_{i<j} m_i^2 m_j^2 ~ q^2 -192\, \Pf m~ q^3 + \ldots~,\\
 \vev{\Tr m^6} & = \Tr \vev{m}^6 + 180\sum_{i<j<k} m_i^2 m_j^2
 m_k^2 ~q^2 + 960\, \Pf m\, \sum_i m_i^2~ q^3 + \ldots ~,\\
 \vev{\Tr m^8} & = \Tr \vev{m}^8 - 420\, (\Pf m)^2 ~q^2 - 2240\,\Pf m\,
 \sum_{i<j}m_i^2 m_j^2~ q^3 + \ldots ~.
\end{aligned}
\end{equation}
It is interesting to observe that the non-perturbative part of these expressions
exactly coincides with the non-perturbative corrections for the coupling constant
of the four-dimensional SU(2) gauge theory with $N_f=4$ as reported in \eq{deltal}.
We will elaborate more on this relation in Section~\ref{sec:intrel}.

\section{Four-dimensional effective prepotential}
\label{sec:4d}
The non-perturbative prepotential for the four-dimensional Sp(1) gauge theory on the D3 branes 
can be computed by extracting the sub-leading $1/(\epsilon_1\epsilon_2)$ divergence in the
deformed D-instanton partition function using formulae (\ref{singlogZ}) and
(\ref{limprep}). 
The result of these calculations, which are technically similar to those
reported in \cite{Billo':2010bd}, can be written as follows
\begin{equation}
 \label{f4is}
  \cF^{(4)}_{\mathrm{n.p.}} = \sum_k \cF_k\, q^k~, 
\end{equation}
where the first five instanton contributions are explicitly given by
 \begin{align}
 \phantom{\Bigg\{}\cF_1&=\frac{2\,\Pf m}{a^2}~,
\label{f21}\\
\phantom{\Bigg\{}\cF_2&=\frac{5\,(\Pf m)^2}{4\,a^6} - \frac{3\sum_{i<j<k}m_i^2 m_j^2
  m_k^2}{2a^4} +\frac{\sum_{i<j} m_i^2 m_j^2}{a^2} 
~,
\label{f22}\\
\phantom{\Bigg\{}\cF_3&=\Pf m\,\Big\{
   \frac{3\,(\Pf m)^2}{a^{10}} -
  \frac{14\sum_{i<j<k} m_i^2 m_j^2 m_k^2} {3\,a^8}
  + \frac{20\sum_{i<j}m_i^2 m_j^2}{3\,a^6}\notag\\ 
&~~~~
-\frac{8\sum_{i}m_i^2}{a^4} + \frac{8}{a^2}\Big\}~,\label{f23}\\
\phantom{\Bigg\{}\mathcal{F}_4 & =
 \frac{1469(\Pf m)^4}{128\,a^{14}} 
 - \frac{715(\Pf m)^2 \sum_{i<j<k}m_i^2 m_j^2 m_k^2}{32\,a^{12}}
 \notag\\ 
 &~~~~ + \frac{153\sum_{i<j<k} m_i^4 m_j^4 m_k^4 + 1332 (\Pf m)^2
\sum_{i<j} m_i^2 m_j^2}{32\,a^{10}}\notag\\ 
 &~~~~ - \frac{63 \sum_i m_i^2 \sum_{j<k\not= i}m_j^4 m_k^4 + 588\, (\Pf
 m)^2 \sum_i m_i^2 }{8\,a^8} \notag\\ 
&~~~~ +\frac{5\sum_{i<j} m_i^4 m_j^4 + 100 \sum_i m_i^4 \sum_{j<k\not=
 i}m_j^2 m_k^2 + 960 \,(\Pf m)^2}{8\,a^6} \notag\\ 
 &~~~~ - \frac{3\sum_{i\not= j}m_i^4 m_j^2 + 36 \sum_{i<j<k}m_i^2 m_j^2
 m_k^2}{2\,a^4}
 + \frac{\sum_i m_i^4 + 4 \sum_{i<j} m_i^2 m_j^2}{2\,a^2}
~,\label{f24}\\
\phantom{\Bigg\{}\mathcal{F}_5 & = \Pf m \Big\{
 \frac{4471\,(\Pf m)^4}{80\,a^{18}} - 
 \frac{525\,(\Pf m)^2\sum_{i<j<k}m_i^2 m_j^2 m_k^2}{4\,a^{16}} \notag\\ 
 &~~~~ + \frac{{1131}\sum_{i<j<k} m_i^4 m_j^4 m_k^4 + 5980\, (\Pf m)^2
 \sum_{i<j} m_i^2 m_j^2}{20\,a^{14}}
 \notag\\ &~~~~
 - \frac{121 \sum_i m_i^2 \sum_{j<k\not= i}m_j^4 m_k^4 + 660 \,(\Pf m)^2
 \sum_i m_i^2 }{a^{12}}
 \notag\\
&~~~~
 + \frac{{207}\sum_{i<j} m_i^4 m_j^4 + 1260 \sum_i m_i^4 \sum_{j<k\not=
 i}m_j^2 m_k^2 
 + 7020 (\Pf m)^2}{5\,a^{10}}
 \notag\\ &~~~~
 - \frac{84 \sum_{i \not= j} m_i^4 m_j^2 + 504 \sum_{i<j<k} m_i^2
 m_j^2 m_k^2}{a^8}
 + \frac{28\sum_i m_i^4 + 160 \sum_{i<j} m_i^2 m_j^2}{a^6} 
 \notag\\ &~~~~
 - \frac{48 \sum_i m_i^2}{a^4} + \frac{12}{a^2}
 \Big\}~.\label{f25}
\end{align}
To the above non-perturbative prepotential we must add the tree-level term
\begin{equation}
 \cF_{\mathrm{tree}}^{(4)} = \frac{\pi\ii\tau_0}{2!}\,\Tr \phi^2 = a^2\,\ln q~,
\label{tree4}
\end{equation}
where in the last step we replaced the adjoint field $\phi$ with its vacuum expectation
value according to (\ref{defa}), and the perturbative 1-loop contribution given
by (see for example \cite{D'Hoker:1999ft})
\begin{equation}
 \begin{aligned}
  \cF_{\mathrm{1-loop}}^{(4)}\! &=
\frac{1}{2}\sum_i\!\Big[
 \big(a + m_i/\sqrt{2}\big)^2\ln\Big(\frac{a + m_i/\sqrt{2}}{\Lambda}\Big)
+\big(a - m_i/\sqrt{2}\big)^2\ln\Big(\frac{a - m_i/\sqrt{2}}{\Lambda}\Big)
\Big]\\
&~~~~-4\,a^2\,\ln\Big(\frac{a}{\Lambda}\Big)~.
 \end{aligned}
\label{1loop4}
\end{equation}
Note that for $N_f=4$ the only dependence on the UV cut-off $\Lambda$ is
in terms proportional to $\sum_i m_i^2 \log (a/\Lambda)$ which, however, do not
contribute to the effective action.

The total prepotential
\begin{equation}
 \cF^{(4)}  = \cF_{\mathrm{tree}}^{(4)} + \cF_{\mathrm{1-loop}}^{(4)} + \cF_{\mathrm{n.p.}}^{(4)}
\label{preptot4}
\end{equation}
determines the effective coupling $\tau$ of the four-dimensional gauge theory on the D3 branes
according to the relation
\begin{equation}
 \label{tauF1}
 2\pi\ii\, \tau = \frac{\partial^2\cF^{(4)}}{\partial a^2}~.
\end{equation}
If we organize the resulting expression as an expansion in inverse powers of $a$, 
we obtain
\begin{eqnarray}
  2\pi\ii\tau & =& 2\ln q - \frac 12 \frac{\sum_i m_i^2}{a^2} + 
  \frac{1}{a^4} \!\left[\!-\frac 18 \sum_i m_i^4 + 12\, \Pf m \,q 
  + 6 \sum_{i<j} m_i^2 m_j^2 ~q^2 + 48 \,\Pf m\,q^3 + \ldots\right] \nonumber\\
  && + \frac{1}{a^6} \left[\!- \frac{1}{24} \sum_i m_i^6 
  - 30 \sum_{i<j<k}m_i^2 m_j^2 m_k^2 ~ q^2 - 160\, \Pf m\, \sum_i m_i^2~ q^3 +
  \ldots \right]\nonumber\\
  &&+ \frac{1}{a^8} \!\left[\!-\frac{1}{64} \sum_i m_i^8 + 
  \frac{105}{2}\, (\Pf m)^2~ q^2
  + 280\, \Pf m\, \sum_{i<j} m_i^2 m_j^2
  ~ q^3 + \ldots\right]+~\ldots~.
\label{taula}
\end{eqnarray}
This completely agrees with the effective coupling reported in \eq{tauswe2} and
derived from the SW curve for the SU(2) $N_f=4$ theory \cite{Seiberg:1994rs}.
Thus, we can conclude that the non-perturbative features predicted by the SW curve
for this theory can be interpreted as D-instanton
effects in the D7/D3 system of type I$^\prime$. Furthermore, this agreement puts on a 
very solid ground the prescription described in Eq. (\ref{singlogZ}), which, as
a matter of fact, was successfully tested already in \cite{Billo':2010bd} in a
different context by exploiting the heterotic/type 
I$^\prime$ duality.

\section{An intriguing relation}
\label{sec:intrel}
\begin{figure}[hbt]
 \begin{center}
\begin{picture}(0,0)%
\includegraphics{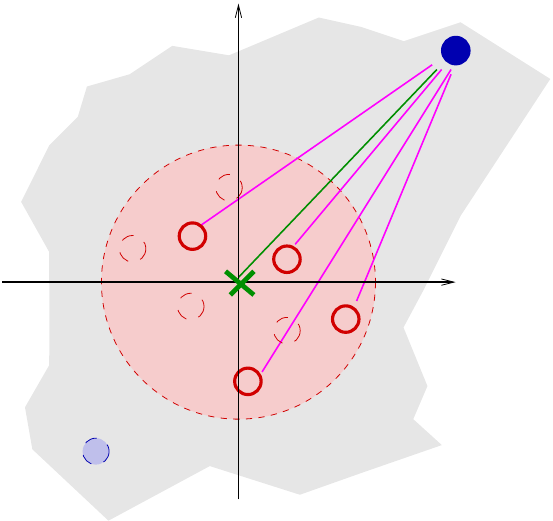}%
\end{picture}%
\setlength{\unitlength}{1989sp}%
\begingroup\makeatletter\ifx\SetFigFont\undefined%
\gdef\SetFigFont#1#2#3#4#5{%
  \reset@font\fontsize{#1}{#2pt}%
  \fontfamily{#3}\fontseries{#4}\fontshape{#5}%
  \selectfont}%
\fi\endgroup%
\begin{picture}(5243,4950)(744,-4359)
\end{picture}%
 \end{center}
 \caption{An ideographic representation of the relation we discuss in this
section (see the main text for a discussion of its content)}
 \label{fig:rel}
\end{figure}

By comparing the non-perturbative terms of the effective coupling (\ref{taula})
with the vacuum expectation values (\ref{chirexp}) of the operators $\Tr
m^{2l}$ forming the chiral ring of the eight-dimensional theory living on the
D7 branes, we find that (up to the available orders in $q$ and in $1/a$) our results
take the following very intriguing and suggestive form
\begin{equation}
 \label{tauswc2}
 {2\pi\ii} \,\tau = {2\pi\ii} \,\tau_0 - \sum_{l=1}^\infty
  \frac{1}{2 l} \frac{\vev{\Tr m^{2l}}}{a^{2l}}~.
\end{equation}
In other words, the \emph{non-perturbative corrections} to the effective
coupling on the D3's are obtained from its \emph{perturbative} expression
(\ref{taupertlf}) after promoting the mass parameters $m_i$, which classically
correspond to the vacuum expectation values of the SO(8) adjoint field $m$ on the D7 branes 
according to \eq{psitom}, to a bona-fide quantum field of which the appropriate expectation
values have to be taken. 
We conjecture%
\footnote{The authors of \cite{Fucito:2009rs} have pushed their computations up
to $\vev{\Tr m^{18}}$ and to instanton number $k=5$, finding that the agreement
with the effective four-dimensional coupling $\tau$ persists \cite{privcomm}.}
that this relation actually extends to all orders
in $q$ and in $1/a^2$, and in Fig.~\ref{fig:rel} we give a graphical
representation of it.
The effective gauge coupling $\tau$ on the probe D3 brane (the blue dot) receives
perturbative contributions.  They arise from loops of open string
states suspended between the D3
and the D7 branes (red circles), their images (dotted red circles) and the
orientifold (green cross). These diagrams have a dual interpretation as
tree-level closed string exchanges which are depicted as straight lines (the
lines to the image D7 branes are not shown to avoid clutter). If we take into
account the full quantum dynamics, including the non-perturbative effects,
of the D7 fields whose eigenvalues correspond to the D7 positions 
(this procedure is graphically indicated by the circular red-shaded area including the D7 branes) 
and we still communicate this information to the probe D3 brane in a perturbative fashion, 
we get the \emph{exact} effective coupling, {\it i.e.} 
\begin{equation}
 \tau_{\mathrm{pert}}~\to~\tau ~~~~~~\mbox{if}~~~~~~\Tr\vev{m}^{2l}~\to~
\vev{\Tr m^{2l}}~.
\label{rel2}
\end{equation}
With simple formal manipulations we can rewrite the relation (\ref{tauswc2}) 
as follows:
\begin{equation}
 a \,\frac{\partial}{\partial a}(2\pi\ii\, \tau) 
= \Big\langle{\Tr \frac{m^2}{a^2-m^2}\Big\rangle} ~,
\label{relnew}
\end{equation}
from which, after using (\ref{tauF1}) in the left-hand side, we obtain
\begin{equation}
\frac{\partial^2}{\partial a^2}\Big( a \,\frac{\partial }{\partial a}\cF^{(4)} -2\cF^{(4)}
\Big) 
= \Big\langle{\Tr \frac{m^2}{a^2-m^2}\Big\rangle} ~.
\label{relnew1}
\end{equation}
Note that $\big(a \,{\partial_a}\cF^{(4)} -2\cF^{(4)}\big)$ is the same combination that
appears in the Matone relation \cite{Matone:1995rx} for the pure SU(2) SYM theory, and hence 
we can regard \eq{relnew1} as a sort of generalization of this relation 
to the SU(2) theory with four massive flavors. 

In conclusion, it can be said that knowing the chiral ring in the
eight-dimensional theory is equivalent to knowing the four-dimensional
prepotential. Conversely the eight-dimensional prepotential can be read off
directly from the four-dimensional one by means of
\begin{equation}
\cF^{(4)}_{\mathrm{n.p.}}= - \frac{1}{a^2} 
\,\Big(q\,\frac{\partial}{\partial_q} \cF^{(8)}_{\mathrm{n.p.}}\Big) + 
{\mathcal O}(\frac{1}{a^4})~.
\end{equation}
This equation is a consequence of the particular relation 
$\frac{1}{4\,!}\vev{\Tr m^{4}}=\,q\,\frac{\partial}{\partial_q} \cF^{(8)}_{\mathrm{n.p.}}$
explained in \cite{Fucito:2009rs}, and of Eq.s (\ref{tauswc2}) and (\ref{tauF1}).

\section{Comparison with the SU(2) instanton calculus}
\label{sec:compare}
In this last section we pursue the comparison of our results with the outcome 
of the standard supersymmetric instanton calculus for the $\mathrm{SU}(2)$ theory with
$N_f$ flavors.

\subsection{Instanton calculus}
\label{subsec:nekra}
The computation of instanton effects in field theories with $\cN=2$
supersymmetry, $\mathrm{SU}(N)$ gauge group and $N_f$ fundamental
hypermultiplets was performed by N. Nekrasov exploiting localization
techniques \cite{Nekrasov:2002qd,Nekrasov:2003af}. 
In fact, it is very easy to extract from his
work explicit expressions for the prepotential, at least up to
three instantons;
for instance, the result for the SU(2) gauge theory with $N_f=3$ is reported here 
in Appendix~\ref{app:dec} (see in particular \eq{Nnf31}). 
This result fully agrees, up to irrelevant constant 
rescalings of the mass parameters and of the dynamical scale, with the limit
in which we decouple one flavor from the four-dimensional prepotential (\ref{preptot4})
obtained for the D3/D7 system in type I$^\prime$. 
This agreement represents another very reassuring check of our procedure.

Although originally proposed for asymptotically free theories, and thus limited
to $N_f<4$ in the case of the $\mathrm{SU}(2)$ gauge theory,
Nekrasov's prescription can formally be extended without any problem also 
to the conformal theory with $N_f=4$. 
We will describe the outcome of Nekrasov's prescription and elaborate on its
relation 
with our results (and the SW approach) in the next subsection by pushing our analysis
up to five instantons.
To do so we resort, 
as a very efficient computational tool, to the recently discovered AGT connection
that allows to obtain the instanton partition function \`a la Nekrasov for the
SU(2) $N_f=4$ gauge theory from the
conformal blocks of a two-dimensional Liouville conformal field theory \cite{Alday:2009aq}.

Before giving some details on this relation, let us anticipate the key points of our findings. 
Denoting by $\tau_{\mathrm{uv}}$
the tree-level coupling in this approach, and by
\begin{equation}
 \label{defquv}
 x = \ee^{\pi\ii\tau_{\mathrm{uv}}}
\end{equation}
the corresponding instanton expansion parameter, one finds
that in the \emph{massless} case 
the \emph{effective} coupling $\tau$ receives instanton corrections and takes the form
\begin{equation}
 \label{teffnek}
\pi\ii\tau=
 \ln x -\ln 16 + 
\frac{1}{2}\,x+\frac{13}{64}\,x^2+\frac{23}{192}\,x^3 + \frac{2701}{32768}\,x^4
+\frac{5057}{81920}\,x^5 +\ldots~. 
\end{equation}
This is to be contrasted with what happens in our treatment and in the SW
description, where for zero masses the effective coupling coincides with the
tree-level coupling $\tau_0$, {\it i.e.}
\begin{equation}
 \label{teffm0}
\pi\ii\tau= \ln q
\end{equation}
as is clear from \eq{taula} when $m_i=0$.
If one assumes that the quantity that should agree in any description of the same
quantum field theory is the effective coupling, then
Eq.s~(\ref{teffnek}) and (\ref{teffm0}) imply
\begin{equation}
\label{relqx}
 \ln q = \ln x -\ln 16 + 
\frac{1}{2}\,x+\frac{13}{64}\,x^2+\frac{23}{192}\,x^3 + \frac{2701}{32768}\,x^4
+\frac{5057}{81920}\,x^5 + \ldots~. 
\end{equation}
One can check that this is the inverse of the expansion given in (\ref{ef8}), so that we
are led to identify $x$ with $\kappa(q)$ and write 
\begin{equation}
 \label{qtoquv}
  x = \frac{\vartheta^4_2(\tau_0)}{\vartheta^4_3(\tau_0)}~,
\end{equation}
which coincides with the closed-form expression proposed in \cite{Alday:2009aq} (see
also \cite{Marshakov:2009gs}--\nocite{Marshakov:2009kj}\cite{Poghossian:2009mk}, and \cite{Grimm:2007tm} where that expression was first proposed). 
As a pure matter-of-fact remark, it is interesting to notice that 
the \emph{tree-level} coupling $\ln x$ utilized in Nekrasov's approach corresponds 
to the \emph{effective} coupling $\ln\kappa(q)$ that appears in front 
of the $\mathrm{SO}(8)$ invariant structure
$T_1$ in the eight-dimensional action on the D7 branes (see Eq.s~(\ref{kappa}) and (\ref{preptot2})).
This is another signal of a non-trivial interplay between quantum effects in four dimensions 
on the D3's and in eight dimensions on the D7's, 
in addition to the one described in Section \ref{sec:intrel}.

Turning to the massive case, one finds that the Nekrasov/AGT prepotential for the
SU(2) $N_f=4$ theory contains, at each order in
its instanton expansion, several terms which do not appear in our findings.
However, if we start from our complete prepotential (\ref{preptot4}) and substitute our 
expansion parameter $q$ in terms of
$x$ according to \eq{relqx}, namely
\begin{equation}
 \label{xtoq}
 q = \frac{x}{16}\Big(1+\frac{1}{2}\,x+\frac{21}{64}\,x^2+\frac{31}{128}\,x^3 
+ \frac{6257}{32768}\,x^4 + \ldots \Big)~,
\end{equation}
we obtain an exact matching with the Nekrasov/AGT prepotential 
(up to constant terms which do not contribute to the effective action).
Let us now show this in some more detail.

\subsection{The Nekrasov prepotential from the AGT realization}
\label{subsec:AGT}
In \cite{Alday:2009aq}
a remarkable relation between
the deformed instanton partition function for $\mathcal N = 2$ SU(2) 
gauge theories in four dimensions
and the correlation functions of the Liouville theory in two dimensions has
been discovered.
In this correspondence the central charge $c$ of the Liouville theory
and the deformation parameters $\epsilon_1$ and $\epsilon_2$ of the instanton
partition functions are related as follows
\begin{equation}
c=1+6\,Q^2=1+ 6\, \frac{(\epsilon_1+\epsilon_2)^2}{\epsilon_1 \epsilon_2}~,
\label{cepsilon}
\end{equation}
while the logarithm of the conformal block of four Liouville operators
\cite{Zamolodchikov:1985ie} is related 
to the non-perturbative prepotential $F_{\mathrm{n.p.}}$
of the $\mathcal N=2$ SU(2) theory with four
massive hypermultiplets as computed in Nekrasov's approach
\cite{Nekrasov:2002qd}--\nocite{Flume:2002az,Bruzzo:2002xf,Nekrasov:2003af}\cite{Nekrasov:2003rj}.
More precisely we have
\begin{equation}
F_{\mathrm{n.p.}}^{(4)} = -\lim_{\epsilon_1,\epsilon_2 \rightarrow 0}\Big\{
{\epsilon_1 \epsilon_2}~
\ln\Big[(1-x)^{2\mu_1\mu_3}
\,B^\Delta_{\Delta_1 \Delta_2; \Delta_3 \Delta_4}(x)
\Big]\Big\}~.
\label{agtrel}
\end{equation}
Here $B^\Delta_{\Delta_1 \Delta_2; \Delta_3 \Delta_4}(x)$ is the conformal block of four Liouville primary fields with conformal dimensions $\Delta_i$, located at
$\infty$, $1$, $x$ and $0$, and factorized
in the channel $(\Delta_1 - \Delta_2)\sim(\Delta_3 - \Delta_4)$ with an
intermediate state of conformal dimension $\Delta$. These dimensions can be
parameterized as follows
 \begin{equation} 
\Delta= \frac{Q^2}{4}-\frac{a^2}{\epsilon_1\epsilon_2}~,~~~~
\Delta_i=\frac{Q^2}{4}-\mu_i^2~,
\label{deltas}
 \end{equation}
where $a$ is the vacuum expectation value of the SU(2) adjoint scalar field, and the
$\mu_i$'s are related to the masses $m_i$ of the fundamental hypermultiplets according to
\begin{equation}
 \begin{aligned}
  \mu_1+\mu_2 +\frac{Q}{2}&=\frac{m_1}{\sqrt{2\epsilon_1\epsilon_2}}~,~~~~
 \mu_1-\mu_2 +\frac{Q}{2}\!\!&=\frac{m_2}{\sqrt{2\epsilon_1\epsilon_2}}~,\\
\mu_3+\mu_4 +\frac{Q}{2}&=\frac{m_3}{\sqrt{2\epsilon_1\epsilon_2}}~,~~~~
 \mu_3-\mu_4 +\frac{Q}{2}\!\!&=\frac{m_4}{\sqrt{2\epsilon_1\epsilon_2}}~.
 \end{aligned}
\label{masses}
\end{equation}
The dressing factor
$(1-x)^{2\mu_1\mu_3}$ has been inserted to decouple a U(1) factor \cite{Alday:2009aq} and 
get the instantonic partition function for SU(2) and not for U(2). 
Note that in the literature there are different expressions 
for such a factor, containing also terms that depend on the background
charge $Q$. However, these terms are not relevant in the limit 
$\epsilon_1,\epsilon_2 \to 0$ we are considering here%
\footnote{These $Q$-dependent pieces are relevant instead if also gravitational corrections
to the gauge theory are considered, {\it i.e.} if the deformations parameters $\epsilon_1$
and $\epsilon_2$ are not switched off.}.

Using the expression of the conformal blocks \cite{Zamolodchikov:1985ie} (see
also \cite{Marshakov:2009gs}--\nocite{Marshakov:2009kj}\cite{Poghossian:2009mk})
and expanding in powers of $x$, we have
\begin{equation}
 \label{f4nek}
  F^{(4)}_{\mathrm{n.p.}} = \sum_k F_k\, x^k~, 
\end{equation}
where the first few coefficients are 
 \begin{align}
  \phantom{\Bigg\{}F_1&=\frac{\Pf m}{8\,a^2}+ \frac{\sum_{i<j} m_i m_j}{4}+\frac{a^2}{2}~,
\label{F1_nek}\\
\phantom{\Bigg\{}F_2&=\frac{5(\Pf m)^2}{1024\,a^6} - \frac{3\sum_{i<j<k}m_i^2 m_j^2
  m_k^2}{512\,a^4} +\frac{\sum_{i<j} m_i^2 m_j^2+16\,\Pf m}{256\,a^2} \notag\\
&~~~~\phantom{\Bigg\{}
+\frac{\sum_i m_i^2+16\sum_{i<j}m_im_j}{128}+\frac{13\,a^2}{64}~,
\label{F2_nek}\\
\phantom{\Bigg\{}F_3&=\frac{3(\Pf m)^3}{4096\,a^{10}}- \frac{7\,\Pf m
\sum_{i<j<k} m_i^2 m_j^2 m_k^2} {6144\,a^8}
+\frac{5\,\Pf m\sum_{i<j}m_i^2 m_j^2+15(\Pf m)^2}{3072\,a^6}\notag\\
&~~~~\phantom{\Bigg\{}
-\frac{3\sum_{i<j<k} m_i^2 m_j^2 m_k^2+\Pf m\sum_{i}m_i^2}{512\,a^4}
+\frac{\sum_{i<j} m_i^2 m_j^2+11\,\Pf m}{256\,a^2}\notag\\
&~~~~\phantom{\Bigg\{}+\frac{3\sum_{i} m_i^2 +32 \sum_{i<j} m_im_j}{384}+\frac{23\,a^2}{192}~.
\label{F3_nek}
 \end{align}
In Appendix~\ref{app:agt} we report also the explicit expressions for $F_4$ and $F_5$ (see
Eq.s (\ref{f4_nek}) and (\ref{f5_nek})), as well as
some other technical details on this approach.

To this non-perturbative prepotential we must add the tree-level contribution
\begin{equation}
 F_{\mathrm{tree}}^{(4)} = a^2\,\ln x~,
\label{tree4_nek}
\end{equation}
and the $x$-independent 1-loop part
\begin{equation}
 \begin{aligned}
  F_{\mathrm{1-loop}}^{(4)}\! &=
\frac{1}{2}\sum_i\!\Big[
 \big(a + m_i/\sqrt{2}\big)^2\ln\Big(\frac{a + m_i/\sqrt{2}}{\Lambda}\Big)
+\big(a - m_i/\sqrt{2}\big)^2\ln\Big(\frac{a - m_i/\sqrt{2}}{\Lambda}\Big)
\Big]\\
&~~~~-4\,a^2\,\ln\Big(\frac{2\,a}{\Lambda}\Big)~,
 \end{aligned}
\label{1loop4_nek}
\end{equation}
which coincides with (\ref{1loop4}) except for a small difference in the last term
where the effect of a finite renormalization pointed out in \cite{Dorey:1996bn} has been taken
into account. 

The total prepotential
\begin{equation}
 F^{(4)}  = F_{\mathrm{tree}}^{(4)} + F_{\mathrm{1-loop}}^{(4)} + F_{\mathrm{n.p.}}^{(4)}
\label{preptot4_nek}
\end{equation}
does not agree, at first sight, with the prepotential $\cF^{(4)}$ we have found
in Section~\ref{sec:4d} (see Eq.s (\ref{f4is})-(\ref{preptot4})). Indeed, one
can easily see that
in the massless case ($m_i=0$) the non-perturbative part
$\cF_{\mathrm{n.p.}}^{(4)}$ is vanishing while $F_{\mathrm{n.p.}}^{(4)}$ is not,
and that the numerical 
coefficients as well as the various structures are different in the two expressions. 
However, one can explicitly check that {\emph{all}} $a$-dependent structures of 
$\cF^{(4)}$ and $F^{(4)}$ are exactly mapped into
each other if $q$ and $x$ are related as in (\ref{xtoq}) (or equivalently as in (\ref{relqx})).
Let us show this in some cases. Consider for example 
the terms in $F^{(4)}$ that are proportional to $a^2$, namely
\begin{equation}
 a^2\,\Big(\ln x -\ln 16 + 
\frac{1}{2}\,x+\frac{13}{64}\,x^2+\frac{23}{192}\,x^3 + \ldots \Big)~.
\label{a2_nek}
\end{equation}
Upon using (\ref{relqx}), they simply become $a^2\,\ln q$, that is exactly the $a^2$ term of
$\cF^{(4)}$ given in (\ref{tree4}). Now consider the terms 
in $F^{(4)}$ that are proportional to $1/a^2$, namely
\begin{equation}
\frac{1}{a^2}\,\Big\{\!-\frac{\sum_im_i^4}{48}+
 \frac{\Pf m}{8}\, x +\frac{\sum_{i<j} m_i^2 m_j^2+16\,\Pf m}{256}\, x^2
+\frac{\sum_{i<j} m_i^2 m_j^2+11\,\Pf m}{256}\,x^3+\ldots\Big\}
\label{a-2_nek}
\end{equation}
where the first $x$-independent contribution arises from $F_{\mathrm{1-loop}}^{(4)}$. If we use
the relation (\ref{xtoq}), we can rewrite (\ref{a-2_nek}) as
\begin{equation}
\frac{1}{a^2}\,\Big\{\!-\frac{\sum_im_i^4}{48}+
 2\,(\Pf m)\, q +\big(\sum_{i<j} m_i^2 m_j^2\big)\, q^2
+8\,(\Pf m)\,q^3+\ldots\Big\}~,
 \label{a-2}
\end{equation}
which exactly coincides with the part of $\cF^{(4)}$ proportional to $1/a^2$ obtained in
Section~\ref{sec:4d}. Likewise one can check that the agreement persists in all other
$a$-dependent terms, and up to five instantons using the expressions reported in Appendix~\ref{app:agt}.

The only mismatch is in the constant $a$-independent part, which however does not contribute to the
four-dimensional effective action obtained from the prepotential by promoting $a$ to
the full-fledged chiral superfield $\Phi$ and integrating over the chiral four-dimensional
superspace. Thus, this mismatch is irrelevant. We also note that while the 
non-perturbative prepotential $\cF_{\mathrm{n.p.}}$ obtained from the D7/D3/D(--1) system is fully 
invariant under the SO(8) flavor symmetry acting on the massive 
multiplets, the prepotential $F_{\mathrm{n.p.}}$ derived above contains the structure 
$\sum_{i<j} m_im_j$ which is not an SO(8) invariant.
Since this structure appears only in the $a$-independent part, it
does not contribute to the effective action, which indeed is SO(8) invariant. 
It is interesting also to notice that all terms proportional to 
$\sum_{i<j} m_im_j$ could be removed from the prepotential by changing the
dressing factor in the AGT relation (\ref{agtrel}), and using $(1-x)^{-\frac{1}{2}(\mu_1^2
-\mu_2^2+\mu_3^2-\mu_4^2)}$ in place of $(1-x)^{2\mu_1\mu_3}$. 
Finally, the $a$-independent
terms proportional to $\sum_i m_i^2$ that were absent in $\cF_{\mathrm{n.p.}}$
could be reabsorbed with
a suitable redefinition of the UV cut-off appearing in the 1-loop part of the prepotential.
As we have already remarked, all these ambiguities in the constant piece
are not significant and do not influence the effective action of the four-dimensional gauge fields.

\section{Summary and conclusions}
\label{sec:concl}
We have considered the local system composed of four D7 branes and a D3 brane
near one O7 fixed plane in type I$^\prime$ theory on a $T_2$ torus. We have computed
by means of localization techniques the non-perturbative corrections due to
D-instantons to the effective actions on the D7 branes (for which they
correspond to exotic instantons) and on the D3 brane. On the D3, which supports
a $\mathrm{Sp}(1)$, $N_f=4$ theory, they represent gauge instantons. The
effective coupling on the D3 coincides with the exact axio-dilaton field, and
indeed we find that it agrees with the F-theoretic prediction put forward long
ago by Sen \cite{Sen:1996vd}. We thus show by an explicit microscopic
computation how F-theory resums D-instanton effects.
Our results are consistent with the outcome of usual instanton calculus \`a la
Nekrasov \cite{Nekrasov:2002qd} 
for the $\mathrm{SU}(2)$, $N_f=4$ conformal theory, or with its AGT
reformulation in terms of 2d Liouville blocks \cite{Alday:2009aq}, upon a
non-perturbative redefinition of the tree-level coupling.

We point out a very interesting relation between the effective coupling  $\tau$
on the D3 (that is, the modular parameter of the F-theory curve) and the
eight-dimensional dynamics on the D7-branes, and specifically the ``chiral
ring'' correlators $\vev{\Tr m^{2l}}$, where $m(X)$ is the eight-dimensional field in the
adjoint of $\mathrm{SO}(8)$ whose eigenvalues $m_i$ appear as mass parameters
on the D3. Indeed, we find that the exact expression for $\tau$ is obtained
from its perturbative part replacing the occurrences of the masses with the
corresponding chiral ring correlators in eight dimensions. 

\vfill \eject

\noindent {\large {\bf Acknowledgments}}
\vskip 0.2cm
We thank R. Argurio, L. Ferro and especially M. Frau, F. Fucito, J.F. Morales 
and R. Poghossian for several useful discussions. M.~B. thanks the K.I.T.P, 
Santa Barbara, and
the organizers of the workshop ``Strings at the LHC and in the Early Universe''
for hospitality during the initial stage of this work. This research
was supported in part by the National Science Foundation under Grant No. NSF
PHY05-51164. 

\vskip 1cm
\appendix

\section{Notations and theta-function conventions}
\label{app:not} 

\paragraph{Couplings:}
We use the following names for the various quadratic gauge couplings:
\begin{itemize}
 \item $\tau_0$ is the tree-level coupling in the D3/D7 system in type I$^\prime$, {\it i.e.}
  the asymptotic value of the axio-dilaton field. In this
  set-up the gauge theory on the D3 is realized as a $\mathrm{Sp}(1)$ theory with four
  flavors.
 \item $\tau_{\rm uv}$ is the tree-level coupling in the usual
  field-theoretical description of the $\mathrm{SU}(2)$ theory with $N_f=4$ flavors.
 \item $\tau$ is the exact \emph{effective} coupling, which depends on the
  tree-level coupling, the vacuum expectation value $a$ of the adjoint scalar parametrizing the
  Coulomb moduli space and on the mass parameters. This coupling must agree in the two
  descriptions. 
\end{itemize}

\paragraph{Theta-functions:}
We adopt the following definition for the Jacobi $\vartheta$-functions:
\begin{equation}
 \label{defthetaf}
 \vartheta[^a_b](v|\tau)=\sum_{n\in \Z} \ee^{\pi\ii\tau(n- \frac a2)^2}
 \ee^{2\pi \ii (n-\frac a2)(v-\frac b2)}~,
\end{equation}
and the usual naming conventions
\begin{equation}
 \label{theta1234}
 \vartheta_1(\tau)=\vartheta[^1_1](0|\tau)~,~~~
 \vartheta_2(\tau)=\vartheta[^1_0](0|\tau)~,~~~
 \vartheta_3(\tau)=\vartheta[^0_0](0|\tau)~,~~~ \vartheta_4(\tau)=\vartheta[^0_1](0|\tau)~. 
\end{equation}
The Dedekind $\eta$-function is defined as
\begin{equation}
 \label{ded}
 \eta(\tau) = \ee^{\frac{\pi\ii\tau}{12}} \prod_{n=1}^\infty 
 \bigl(1-\ee^{2\pi\ii\tau\,n}\bigr)~.
\end{equation}
The following properties are relevant for us:
\begin{align}
 \label{thetaprops}
  \vartheta_3^4(\tau) - \vartheta_4^4(\tau) - \vartheta_2^4(\tau) = 0~,~~~
  \vartheta_2(\tau)\, \vartheta_3(\tau)\, \vartheta_4(\tau)  = 2\,\eta^3(\tau)~. 
\end{align}
We will also use the transformation properties under the
modular group generators, which are

\begin{equation}
 \label{Smodprop}
 S: \left\{
  \begin{aligned} 
   \vartheta[^a_b](v/\tau|- 1/\tau)
   & = \sqrt{-\ii\tau}\,\ee^ {\frac{\pi\ii}{2}(ab+
   \frac{2v^2}{\tau})} \, \vartheta[^b_{-a}](v|\tau)~,\\
   \eta(-1/\tau) & = \sqrt{-\ii\tau} \,\eta(\tau)\phantom{\Big|}~,
  \end{aligned} 
 \right.
\end{equation}
and

\begin{equation}
 \label{Tmodprop}
 \hspace{-45pt}T: \left\{
  \begin{aligned}
   \vartheta[^a_b](v|\tau+1) & =\ee^{-\frac{\pi\ii}{4}a(a-2) } 
   \vartheta[^a_{a+b}](v|\tau)~,\\
   \eta(\tau+1) & = \ee^{\frac{\pi\ii}{12}}\eta(\tau)~.
  \end{aligned} 
 \right.
\end{equation}

\section{Flavor invariants}
\label{app:flavinv}
The parameters $m_i$ correspond to the components along the Cartan directions
of Spin$(8)$ of the vacuum expectation values of the D7/D7 scalar field $m(x)$ introduced
in \eq{psitom}.
In a given representation $\cR$, the element $m^i H_i$ of the algebra is
diagonal in the basis of weights ${\vec w}_A$ ($A=1,\ldots {\rm dim}\, \cR$),
with eigenvalues $\vec m \cdot {\vec w}_A$. 

Spin$(8)$ has an $S_3$ group of external automorphisms (``triality'') that
permutes its three 8-dimensional representations: the vector ($v$), chiral
spinor ($s$) and antichiral spinor ($c$), remapping the weights of each
of these representations in those of another one. This $S_3$ can be generated by
any two exchanges, say
\begin{equation}
 \label{g1g2}
  g_1: ~ s\leftrightarrow c~,~~~
  g_2: ~ v\leftrightarrow c~.
\end{equation}
Given the form of the weights of these representations, we can view these
triality operations as acting on the Cartan parameters $m_i$, as follows:
\begin{equation}
 \label{trial}
 \begin{aligned}
  g_1:~& \{m_1,\,m_2,\,m_3,\,m_4\} \to \{m_1,\,m_2,\,m_3,\,-m_4\}~,\\
  g_2:~& \{m_1,\,m_2,\,m_3,\,m_4\} \to \frac 12
  \{m_1 + m_2 + m_3 - m_4,\,\,m_1 + m_2 - m_3 + m_4,\\
  & \hskip 4.05cm  m_1 - m_2 + m_3 + m_4,\,\,-m_1 + m_2 + m_3 + m_4\}~.
 \end{aligned}
\end{equation}

The SW curve for the SU(2) theory with $N_f=4$ flavors, 
described in Section~\ref{subsubsec:swc}, is written
in terms of SO$(8)$-invariant expressions of order 2, 4 and 6 in the masses
$m_i$ \cite{Seiberg:1994aj}. The quadratic invariant is
\begin{equation}
 \label{Rdef}
 R = \frac 12 \sum_i m_i^2 = \frac 12 \Tr m^2 ~,
\end{equation}
where the second equality follows from (\ref{psitom}).
This expression is also invariant under triality, as one can see by applying
to it the generators (\ref{trial}). 

The three independent quartic SO(8) invariants can be chosen to be $R^2$ and 
\begin{equation}
 \label{Tdef}
 \begin{aligned}
 T_1 & = \frac{1}{12} \sum_{i<j} m_i^2 m_j^2 - \frac{1}{24} \sum_i m_i^4
=\frac{1}{24}\left(\Tr m^2\right)^2 - \frac{1}{6}\,\Tr m^4~,\\
 T_2 & = -\frac{1}{24} \sum_{i<j} m_i^2 m_j^2 + \frac{1}{48} \sum_i m_i^4 
-\frac 12 \Pf m=-\frac{1}{48}\left(\Tr m^2\right)^2 + \frac{1}{12}\,\Tr m^4
-\frac 12 \Pf m~.
 \end{aligned}
\end{equation}
The triality generators act on the invariants $T_\ell$ as follows:
\begin{equation}
 \label{triT}
  g_1:~ T_2\longleftrightarrow T_3=-T_1-T_2~~~\mbox{and}~~~
  g_2:~ T_1\longleftrightarrow T_2~.
\end{equation}
In other words, the triality group $S_3$ permutes the $T_\ell$ invariants. 

The four sextic SO(8) invariants can be chosen to be $R^3$, $R T_\ell$ and
\begin{equation}
 \label{Ndef}
 N = \frac{3}{16} \sum_{i<j<k} m_i^2 m_j^2 m_k^2 - \frac{1}{96} 
 \sum_{i\not= j} m_i^2 m_j^4 + \frac{1}{96} \sum_i m_i^6~
\end{equation}
which is invariant also under triality.

The following combinations are relevant for us:
\begin{equation}
 \label{combinv}
 \begin{aligned} 
&  R^2 + 6\, T_1  = \sum_{i<j} m_i^2 m_j^2~,\\
& R^2 - 6\, T_1  = \frac 12 \sum_i m_i^4~,\\
 & 2 \,N + R \,T_1  = \frac 12 \sum_{i<j<k} m_i^2 m_j^2 m_k^2~,\\
 & 6\,N + R^3 - 15\, R\, T_1 = \frac 12 \sum_i m_i^6~,\\
 & 2 (\Pf m)^2 - 16\, R\, N - R^4 + 28 \,R^2 \,T_1 - 36 \,T_1^2  = -\frac 12 \sum_i m_i^8~.
 \end{aligned}
\end{equation}

\subsection{Triality extension of SL$(2,\mathbb{Z})$}
\label{subsec:trsl}
Consider an SL$(2,\mathbb{Z})$ group generated by $\cS$ and $\cT$ subject to the
relations
\begin{equation}
 \label{STgen}
  \cS^2 = 1~,~~~ (\cS\cT)^3 = 1~. 
\end{equation}
In our case this group is the modular group
acting on the bare coupling $\tau_0$. Now consider also the permutation group $S_3$ generated
by $g_1$ and $g_2$, subject to the relations
\begin{equation}
 \label{ggen}
 g_1^2 = 1~,~~~ g_2^2 = 1~,~~~ (g_2 g_1)^3 = 1~.
\end{equation}
In our case this group is the triality group of SO(8), we have described above.

Let us then construct the semi-direct product of SL$(2,\mathbb{Z})$ and $S_3$,
which is generated by the following combined actions:
\begin{equation}
 \label{STtilde}
  \widehat\cT = (\cT,\,g_1)~,~~~
  \widehat\cS = (\cS,\,g_2)~.
\end{equation}
These two generators obey
\begin{equation}
 \label{STtrel}
 \widehat\cS^{\,2} = 1~,~~~ (\widehat\cS\,\widehat\cT)^3 = 1~;
\end{equation}
therefore this group is again isomorphic to SL$(2,\mathbb{Z})$.

\section{Details on the D-instanton computation}
\label{app:details}
The expressions of the factors $P_{\Yasymm}(\chi)$, $P_{\Ysymm}(\chi)$
and $P_{\Yfund}(\chi)$ appearing in the integrand of the formula (\ref{zkloc}) for
the instanton partition function are given respectively in Eq.s~(\ref{padj}), (\ref{psym})
and (\ref{Pfun}) in terms of the weights of the relevant
representations of the SO$(k)$ instantonic symmetry group. Let us recall the
form of these weight vectors.
\paragraph{Weight sets of SO$(2n+1)$}
This group has rank $n$.
If we denote by $\ve{i}$ the versors in the 
$\mathbb{R}^n$ weight space, then
\begin{itemize}
 \item the set of the $2n+1$ weights $\vec\pi$ of the vector representation is 
 given by
 \begin{equation}
  \label{setvec1}
  \pm\ve i~,~~~~ \vec 0~~\mbox{with multiplicity $1$}~;
 \end{equation}
 \item the set of $n(2n+1)$ weights $\vec\rho$ of the adjoint representation 
(corresponding to the two-index antisymmetric tensor) is the following:
\begin{equation}
 \label{setrho1}
 \pm \ve i \pm \ve j~(i < j)~,~~~~
 \pm \ve i~,~~~~
 \vec 0~~\mbox{with multiplicity $n$}~;
\end{equation}
\item the $(n+1)(2n+1)$ weights of the two-index symmetric tensor%
\footnote{In fact, this is not an irreducible representation: it decomposes 
into the $(n+1)(2n+1)-1$ traceless symmetric tensor plus a singlet. One of the
$\vec 0$ weights corresponds to the singlet.} are
\begin{equation}
 \label{setsymm1}
 \pm \ve i \pm \ve j~(i < j)~,~~~~
 \pm \ve i~,~~~~
 \pm 2 \ve i~,~~~~
 \vec 0~~\mbox{with multiplicity $n+1$}~.
\end{equation} 
\end{itemize}

\paragraph{Weight sets of SO$(2n)$}
This group has rank $n$.
If we denote by $\ve{i}$ the versors in the 
$\mathbb{R}^n$ weight space, then
\begin{itemize}
 \item the set of the $2n$ weights $\vec\pi$ of the vector representation is 
 given by
 \begin{equation}
  \label{setvec}
  \pm\ve i~;
 \end{equation}
 \item the set of $n(2n-1)$ weights $\vec\rho$ of the adjoint representation 
(corresponding to the two-index antisymmetric tensor) is the following:
\begin{equation}
 \label{setrho}
 \pm \ve i \pm \ve j~(i < j)~,~~~~
 \vec 0~~\mbox{with multiplicity $n$}~;
\end{equation}
\item the $n(2n+1)$ weights of the two-index symmetric tensor%
\footnote{Again, this is not an irreducible representation, since it 
contains a singlet.} are
\begin{equation}
 \label{setsymm}
 \pm \ve i \pm \ve j~(i < j)~,~~~~
 \pm 2 \ve i~,~~~~
 \vec 0~~\mbox{with multiplicity $n$}~.
\end{equation}
\end{itemize}
Inserting these expressions into \eq{padj} we have
\begin{equation}
 \label{padjeo}
 P_{\Yasymm} = \left\{
 \begin{aligned}
 & (- s_1 s_2 s_3)^n  \prod_{l=1}^3 \prod_{I>J=1}^n 
 \bigl[(\chi_I - \chi_J)^2 - s_l^2 \bigr]
 \bigl[(\chi_I + \chi_J)^2 - s_l^2 \bigr] & \mbox{for } k=2n~,\\
 & (- s_1 s_2 s_3)^n  \prod_{l=1}^3 \prod_{I>J=1}^n 
 \bigl[(\chi_I - \chi_J)^2 - s_l^2 \bigr]
 \bigl[(\chi_I + \chi_J)^2 - s_l^2 \bigr] &\null \\
 & \phantom{(- s_1 s_2 s_3)^n}\hskip -8 pt\times 
 \prod_{K=1}^n (\chi_K^2 - s_l^2) & \mbox{for } k=2n+1~.
 \end{aligned}
 \right.
\end{equation}
{From} \eq{psym} we get instead
\begin{equation}
 \label{psymeo}
 P_{\Ysymm} = \left\{
 \begin{aligned}
 & (\epsilon_1\epsilon_2\epsilon_3\epsilon_4))^n  \prod_{A=1}^4 
 \prod_{I>J=1}^n 
 \bigl[(\chi_I - \chi_J)^2 - \epsilon_A^2 \bigr]
 \bigl[(\chi_I + \chi_J)^2 - \epsilon_A^2 \bigr] & \null\\
 & \phantom{(\epsilon_1\epsilon_2\epsilon_3\epsilon_4))^{n+1}}\hskip -8 pt
  \times \prod_{K=1}^n (4\chi_K^2 - \epsilon_A^2)
 & \hskip -115 pt\mbox{for } k=2n~,\\
 & (\epsilon_1\epsilon_2\epsilon_3\epsilon_4))^{n+1}  \prod_{A=1}^4 
 \prod_{I>J=1}^n 
 \bigl[(\chi_I - \chi_J)^2 - \epsilon_A^2 \bigr]
 \bigl[(\chi_I + \chi_J)^2 - \epsilon_A^2 \bigr] & \null\\
 & \phantom{(\epsilon_1\epsilon_2\epsilon_3\epsilon_4))^{n+1}}\hskip -8 pt
  \times \prod_{K=1}^n (4\chi_K^2 - \epsilon_A^2) (\chi_K^2 - \epsilon_A^2)
 & \hskip -75 pt\mbox{for } k=2n+1~. 
 \end{aligned}
 \right.
\end{equation}
{From} \eq{Pfun} we find
\begin{equation}
 \label{Pfune}
  P_{\Yfund} = \prod_{I=1}^n \prod_{i=1}^4 \bigl(\chi_I^2 +
  \frac{m_i^2}{2}\bigr)
  \frac{\bigl[\chi_I^2 - 
  \bigl(\ii a + \frac{\epsilon_3 - \epsilon_4}{2}\bigr)^2\bigr]
  \bigl[\chi_I^2 - 
  \bigl(\ii a - \frac{\epsilon_3 - \epsilon_4}{2}\bigr)^2\bigr]}{\bigl[\chi_I^2
  - \bigl(\ii a + \frac{\epsilon_1 + \epsilon_2}{2} \bigr)^2\bigr]
  \bigl[\chi_I^2 - 
  \bigl(\ii a - \frac{\epsilon_1 + \epsilon_2}{2}\bigr)^2\bigr]}~~~
  \mbox{for } k = 2n~,
\end{equation}
while for $k$ odd we get simply an extra null-vector contribution:
\begin{equation}
 \label{Pfuno}
 P_{\Yfund} = \frac{\Pf m}{4}\, \frac{a^2 - (\frac{\epsilon_3 -
  \epsilon_4}{2})^2}{a^2 - (\frac{\epsilon_1  + \epsilon_2}{2})^2} \times
  \mbox{r.h.s. of \eq{Pfune}}~~~\mbox{for } k = 2n+1~.
\end{equation}
Finally, the Vandermonde determinant $\Delta(\chi)$ is explicitly given by
\begin{equation}
 \label{vande}
 \Delta = \left\{
 \begin{aligned}
 & \prod_{I>J=1}^n (\chi_I^2 - \chi_J^2)^2 
 & ~~\mbox{for } k=2n~,\\
 & \prod_{I>J=1}^n (\chi_I^2 - \chi_J^2)^2
   \prod_{K=1}^n \chi_K^2
 & ~~~~\mbox{for } k=2n+1~. 
 \end{aligned}
 \right.
\end{equation}

\section{Details on the AGT correspondence}
\label{app:agt}

The four-point conformal block $B^\Delta_{\Delta_1 \Delta_2; \Delta_3 \Delta_4}(x)$
appearing in the AGT relation (\ref{agtrel}) can be written \cite{Zamolodchikov:1985ie}
as a formal series according to
\begin{equation}
B^\Delta_{\Delta_1 \Delta_2; \Delta_3 \Delta_4}(x)
=\sum_{|Y|=|Y'|} x^{|Y|} ~\gamma_{\Delta_1\Delta_2\Delta}(Y)~Q_\Delta^{-1}(Y,Y')
~\gamma_{\Delta_3\Delta_4\Delta}(Y')~.
\label{bx}
\end{equation}
Here $Y=\{n_1\geq n_2\geq \cdots\geq n_l>0\}$ is a partition, $|Y|=n_1+n_2+\cdots+n_l$ is its
order, $Q_\Delta(Y,Y')$ is the correlator of two descendants of a primary state
of dimension $\Delta$, namely
\begin{equation}
Q_\Delta(Y,Y')
= \langle\Delta |\, \cL_Y\, \cL_{-Y'} |\Delta \rangle~,
\label{qx}
\end{equation}
where
\begin{equation}
 \cL_{-Y} \,|\Delta \rangle = {L}_{-n_1}{L}_{-n_2}\cdots{L}_{-n_l}\,|\Delta \rangle
\label{ldelta}
\end{equation}
with $L_n$ being the Virasoro generators, and finally $\gamma_{\Delta_1\Delta_2\Delta}(Y)$ 
is given by 
\begin{equation}
\gamma_{\Delta_1\Delta_2\Delta}(Y)
=\prod_{l}\big(\Delta+n_l\,\Delta_1-\Delta_2+\sum_{r<l}n_r\big)~.
\label{gammadeltas}
\end{equation}
For further details, see for example \cite{Marshakov:2009gs}.

The explicit evaluation of $Q_\Delta(Y,Y')$ and its inverse is a straightforward but tedious 
task. To perform the algebra we used the
open-source program REDUCE and, exploiting the AGT relation (\ref{agtrel}), were
able to compute%
\footnote{We found convenient to compute the conformal blocks in the particular
case $\epsilon_1 + \epsilon_2 = 0$, {\it i.e.} for $Q=0$ and $c=1$; this has no
implications on the result, since according to \eq{agtrel} in the computation of
$F_{\mathrm{n.p.}}^{(4)}$ only the limit in which the
$\epsilon_{1,2}$ parameters vanish matter. Here we would like comment that,
despite the fact that the only unitary $c=1$ conformal theory is the free one,
the conformal block we computed is not the one associated with the free theory.
This happens because the conformal block is derived using the Virasoro algebra only,
and thus the conformal dimension of the intermediate state is arbitrary, which would
not happen in the free theory.} the non-perturbative prepotential
$F_{\mathrm{n.p.}}^{(4)}$ up to the fifth order in $x$. The first three
coefficients $F_1$, $F_2$ and $F_3$ were reported in the main text in Eq.s
(\ref{F1_nek}), (\ref{F2_nek}) and (\ref{F3_nek}) respectively. Here we give the
expressions for the fourth and fifth coefficients. We have
\begin{align}
 \phantom{\Bigg\{}{F}_4 & =
 \frac{1469(\Pf m)^4}{8388608\,a^{14}} 
 - \frac{715(\Pf m)^2 \sum_{i<j<k}m_i^2 m_j^2 m_k^2}{2097152\,a^{12}}
 \notag\\ 
&\phantom{\Bigg\{} + \frac{153\sum_{i<j<k} m_i^4 m_j^4 m_k^4 + 1332 (\Pf m)^2
 \sum_{i<j} m_i^2 m_j^2+2304 (\Pf m)^3}{2097152\,a^{10}}
 \notag\\ 
&\phantom{\Bigg\{} - \frac{63 \sum_i m_i^2 \sum_{j<k\not= i}m_j^4 m_k^4 + 588\, (\Pf
 m)^2 \sum_i m_i^2+896\,\Pf m \sum_{i<j<k}m_i^2 m_j^2 m_k^2 }{524288\,a^8}
 \notag\\ 
&\phantom{\Bigg\{}+\frac{5\sum_{i<j} m_i^4 m_j^4 + 100 \sum_i m_i^4 
\sum_{j<k\not=i}m_j^2 m_k^2 + 3280 \,(\Pf m)^2+1280\,\Pf m \sum_{i<j}m_i^2 m_j^2}{524288\,a^6}
 \notag\\ 
&\phantom{\Bigg\{}-\frac{3\sum_{i\not= j}m_i^4 m_j^2 + 732 \sum_{i<j<k}m_i^2 m_j^2
 m_k^2+384\,\Pf m \sum_{i}m_i^2}{131072\,a^4}\notag\\
&\phantom{\Bigg\{}+ \frac{\sum_i m_i^4 + 468\sum_{i<j} m_i^2 m_j^2+4352\,\Pf m}{131072\,a^2}
 \notag\\
&\phantom{\Bigg\{}+ {\frac{233\sum_i m_i^2+2048\sum_{i<j}m_im_j}{32768}}+
\frac{2701\,a^2}{32768}
\label{f4_nek}
\end{align}
and
\begin{align}
 \phantom{\Bigg\{}{F}_5 & = \frac{4471(\Pf m)^5}{83886080 \,a^{18}} - 
 \frac{525(\Pf m)^3 \sum_{i<j<k}m_i^2 m_j^2 m_k^2}{4194304\,a^{16}} \notag\\
&\phantom{\Bigg\{}  + \frac{1131\,\Pf m\,\sum_{i<j<k} m_i^4 m_j^4 m_k^4 
+5980(\Pf m)^3 \sum_{i<j} m_i^2 m_j^2 + 7345(\Pf m)^4
}{20971520\,a^{14}}
  \notag\\ 
&\phantom{\Bigg\{}-\frac{121\Pf m \sum_i m_i^2 \sum_{j<k\not= i}m_j^4 m_k^4
+715(\Pf m)^2\! \sum_{i<j<k} m_i^2 m_j^2 m_k^2+
660(\Pf m)^3\! \sum_i m_i^2 } {10 48 576\,a^{12}}
\notag \\
&\phantom{\Bigg\{}+\frac{1}{5242880\,a^{10}}\Bigg[765\sum_{i<j<k} m_i^4 m_j^4 m_k^4+1260\,\Pf m\sum_i m_i^4 \sum_{j<k\not= i}m_j^2 m_k^2\notag\\
&\phantom{\Bigg\{+\frac{1}{5242880~~a^{10}}}
+207\,\Pf m \sum_{i<j} m_i^4 m_j^4+6660(\Pf m)^2 \sum_{i<j} m_i^2 m_j^2+13680(\Pf m)^3\Bigg]
\notag\\
&\phantom{\Bigg\{}-\frac{1}{262144\,a^{8}}\Bigg[63\sum_i m_i^2 \sum_{j<k\not= i}m_j^4 m_k^4 
+644\,\Pf m \sum_{i<j<k} m_i^2 m_j^2 m_k^2
\notag\\
&\phantom{\Bigg\{-\frac{1}{262144\,a^{8}}}~+
21\,\Pf m \sum_{i \not= j} m_i^4 m_j^2 +588(\Pf m)^2 \sum_i m_i^2 \Bigg]\notag\\
&\phantom{\Bigg\{}+\frac{1}{262144\,a^{6}}\Bigg[5\sum_{i<j} m_i^4 m_j^4+100
\sum_i m_i^4 \sum_{j<k\not= i}m_j^2 m_k^2 + 7\,\Pf m \sum_i m_i^4
\notag\\
&\phantom{\Bigg\{\Bigg\{+\frac{1}{262144\,a^{6}}}~~+780\,\Pf m \sum_{i<j} m_i^2 m_j^2
+2000(\Pf m)^2\Bigg]\notag
\\
&\phantom{\Bigg\{}-\frac{3\sum_i m_i^4 \sum_{j\not= i}m_j^2
+348\sum_{i<j<k} m_i^2 m_j^2 m_k^2 +225\,\Pf m \sum_i m_i^2}{65 536\,a^4} 
\phantom{\Bigg\{+\frac{1}{262144\,a^{6}}}~~~~~~
\notag\\
&\phantom{\Bigg\{}+\frac{\sum_i m_i^4+ 212\sum_{i<j} m_i^2 m_j^2+1787\,\Pf
m}{65536\,a^2}
\notag\\
&\phantom{\Bigg\{}+\frac{525\sum_i m_i^2+4096\sum_{i<j}m_i m_j}{81920}
+ \frac{5057\,a^2}{81920} ~.
\label{f5_nek}
\end{align}

\section{Decoupling limits to $N_f=3,2,1,0$.}
\label{app:dec}
{From} the prepotential we derived in Eq.s (\ref{f21})--(\ref{f25}) for the $N_f=4$
case, it is possible to deduce the prepotential for the asymptotically free theories with
$N_f=3,2,1,0$ by decoupling in turn the massive fundamental hypermultiplets.
We will now compare the results obtained in this way with those following from
Nekrasov's prescriptions for the instanton calculus, which were
originally given just for these asymptotically free cases. We will find full
agreement in all terms contributing to the effective action. This enhances our
confidence in our results, and shows that the difference with Nekrasov's
formulas (or, equivalently, with the AGT methods) discussed in Section~\ref{sec:compare} 
only occurs in the conformal case $N_f=4$.

Let us start decoupling one flavor and get the $N_f=3$ theory by taking the limit
\begin{equation}
 \label{m4lim}
 m_4\to\infty~~~~\mbox{with }\, \Lambda\equiv  q\,m_4~~\mbox{fixed}~. 
\end{equation}
{From} Eq.s (\ref{f21})--(\ref{f23}) we find
\begin{eqnarray}
  \cF_{\mathrm{n.p.}}^{(4)}
&= &\frac{2\,m_1 m_2 m_3}{a^2}\,\Lambda
 +\left\{\frac{5(m_1 m_2 m_3)^2}{4\,a^6}
 -\frac{3\sum_{i<j=1}^3 m_i^2 m_j^2}{2\,a^4} + 
 \frac{\sum_{i=1}^3 m_i^2}{a^2} + \frac 14\right\}\!\Lambda^2 
\label{nf3}\\
  && + m_1 m_2 m_3\left\{\frac{3(m_1 m_2 m_3)^2}{a^{10}} -
 \frac{14\sum_{i<j=1}^3 m_i^2 m_j^2}{3\,a^8} +
 \frac{20\sum_{i=1}^3 m_i^2}{3\,a^6} - \frac{8}{a^4}
 \right\} \!\Lambda^3 + O(\Lambda^4)~.
\nonumber
\end{eqnarray}
On the other hand, from Nekrasov's paper \cite{Nekrasov:2003af}, using Eq. (3.16) and the relevant
definitions given in the previous pages therein, we can extract the
following expression of the SU(2) prepotential for $N_f=3$ flavors with
masses $M_i$:
\begin{eqnarray}
 {F}_{\mathrm{n.p.}} 
 &= &\frac{M_1 M_2 M_3+ a^2(M_1 + M_2 + M_3) }{2\, a^2} \,\widehat\Lambda \nonumber\\
 && +\ \frac{5 M_1^2 M_2^2 M_3^2 - 3 a^2 \sum_{i<j=1}^3 M_i^2 M_j^2
 + a^4\sum_{i=1}^3 M_i^2  + a^6}{64 \,a^6} \,{\widehat\Lambda}^2
\label{Nnf31}\\
 && + \,M_1 M_2 M_3 \frac{9 M_1^2 M_2^2 M_3^2 - 7 a^2 \sum_{i<j=1}^3 M_i^2 M_j^2 
    + 5 a^4 \sum_{i=1}^3 M_i^2 - 3 a^6}{192\, a^{10}}
    \,{\widehat\Lambda}^3 + O({\widehat\Lambda}^4)~.
\nonumber
\end{eqnarray} 
The two expressions (\ref{nf3}) and (\ref{Nnf31}) match after taking into account 
a different normalization of the masses and of the dynamical scale, namely
\begin{equation}
 \label{mtoM}
 m_i = \sqrt{2}\, M_i~,~~~
 \Lambda = \frac{\widehat\Lambda}{8\sqrt{2}}~.
\end{equation}
The only difference is the constant term proportional to $(M_1 + M_2 + M_3)$ in
the first line of (\ref{Nnf31}), which however does not contribute to the effective
action upon integration over the chiral superspace, and thus is irrelevant.

The agreement persists for lower values of $N_f$, upon taking further
decoupling limits. For instance, the pure SU(2) case ($N_f=0$) can be reached
from our $N_f=4$ results in Eq.s (\ref{f21})--(\ref{f23}) by sending
\begin{equation}
 \label{nf0limit}
  m_{1,2,3,4}\to \infty~,~~~\mbox{with }\, \Lambda^4\equiv q\,\Pf m
  ~~\mbox{fixed}~.
\end{equation}
The resulting prepotential reads
\begin{equation}
 \label{nf0p}
 \cF = 2 \frac{\Lambda^4}{a^2} + \frac 54 \frac{\Lambda^8}{a^6} + 3
\frac{\Lambda^{12}}{a^{10}} + \frac{1469}{128} \frac{\Lambda^{16}}{a^{14}}
+ \frac{4471}{80}\frac{\Lambda^{20}}{a^{18}} + \ldots
\end{equation}
and coincides with Nekrasov's result if we set $\Lambda^4 = {\hat\Lambda}^4/4$.
The above results are perfectly consistent also with the instanton expansion of the
effective coupling derived from the SW curves
\cite{Seiberg:1994rs,Seiberg:1994aj} for SU(2) gauge theories with $N_f=0,1,2,3$. 

Also the relation (\ref{tauswc2}) between the exact gauge coupling $\tau$ on the D3 branes
and the eight-dimensional chiral ring for the conformal
$N_f=4$ case can be extended to the non-conformal cases with lower $N_f$.
We can check this by exploiting the results of \cite{Fucito:2009rs}, where the first chiral ring
expectation values $\vev{\Tr m^{2l}}$ have been given in the generic
case of $N_f$ D7-branes supporting an SO$(2N_f)$ group. Therefore we can compare
these expressions to the effective coupling derived from the quadratic prepotentials we
just described. Let us consider as an example the $N_f=3$ case. Deriving the prepotential
(\ref{nf3}) with respect to $a$ or, equivalently, taking the limit (\ref{m4lim}) on the
non-perturbative effective coupling (\ref{taula}), we find
\begin{equation}
 \label{taula3}
 \begin{aligned}
  2\pi\ii\,\tau_{\mathrm{n.p.}} & = \frac{1}{a^4} \Bigl(12\, m_1 m_2 m_3\, \Lambda + 
 6 \sum_{i=1}^3 m_i^2\, \Lambda^2 + \ldots\Bigr)\\
  & - \frac{1}{a^6}\Bigl(30 \sum_{i<j=1}^3 m_i^2 m_j^2\, \Lambda^2
  + 160\, m_1 m_2 m_3 \, \Lambda^3 + \ldots\Bigr)\\
  & + \frac{1}{a^8} \Bigl(\frac{105}{2} (m_1 m_2 m_3)^2\, \Lambda^2 +
  280\,m_1 m_2 m_3 \sum_{i=1}^3 m_i^2 \, \Lambda^3 + \ldots\Bigr) +
  \ldots~.
 \end{aligned}
\end{equation}
On the other hand, from \cite{Fucito:2009rs} we can deduce, in our conventions, 
the following results for $N_f=3$
\begin{equation}
\begin{aligned}
 \label{chirexp3}
 \vev{\Tr m^2} & = \Tr \vev{m}^2\phantom{\Big(}~,\\
 \vev{\Tr m^4} & = \Tr \vev{m}^4 - 48\, m_1 m_2 m_3 \, \Lambda
-24\sum_{i=1}^3 m_i^2 \, \Lambda^2 + \ldots~,\\
 \vev{\Tr m^6} & = \Tr \vev{m}^6 + 180\sum_{i<j=1}^3 m_i^2 m_j^2
 \, \Lambda^2  + 960\, m_1 m_2 m_3\, \Lambda^3 + \ldots ~,\\
 \vev{\Tr m^8} & = \Tr \vev{m}^8 - 420\, (m_1 m_2 m_3)^2\, \Lambda^2 
- 2240\,m_1 m_2 m_3 \, \sum_{i=1}^3 m_i^2\, \Lambda^3 + \ldots~.
\end{aligned}
\end{equation}
We therefore see that also in this asymptotically free case the lowest orders
in $\Lambda $ and in $1/a$ are compatible with the relation 
\begin{equation}
 \label{tnpe}
  2\pi\ii\,\tau_{\mathrm{n.p.}}  = -\sum_{l=1}^\infty \frac{1}{2l}
  \frac{\vev{\Tr\, m^{2l}}\big|_{\mathrm{n.p.}} }{a^{2l}}~,
\end{equation}
from which \eq{tauswc2} easily follows. The full effective coupling
$\tau$ can thus be obtained from its perturbative expression (\ref{taupertlf})
upon taking into account the eight-dimensional quantum dynamics of the mass
parameters $m_i$.

The chiral ring is mathematically defined for any number $N_f$ of D7 branes but,
as emphasized in \cite{Fucito:2009rs}, only for $N_f > 4$ (resp. $N_f  = 4$ ) it
can be associated to an eight-dimensional $\mathrm{SO}(2N_f)$ gauge theory with
negative (resp. vanishing) $\beta$ function for which a weak coupling analysis
is reliable. Our analysis in the cases $N_f=4$ and $N_f< 4$ shows that the
chiral ring for those values of $N_f$ can be given a sensible meaning in the
four-dimensional theory.

\providecommand{\href}[2]{#2}\begingroup\raggedright
\endgroup


\begin{thebibliography}{10}

\bibitem{Blumenhagen:2005mu}
R.~Blumenhagen, M.~Cvetic, P.~Langacker, and G.~Shiu, \emph{{Toward realistic
  intersecting D-brane models}}, Ann. Rev. Nucl. Part. Sci. {\bf 55} (2005)
  71--139,
\href{http://arxiv.org/abs/hep-th/0502005}{{\tt arXiv:hep-th/0502005}}.

\bibitem{Blumenhagen:2006ci}
R.~Blumenhagen, B.~Kors, D.~Lust, and S.~Stieberger, \emph{{Four-dimensional
  String Compactifications with D-Branes, Orientifolds and Fluxes}},
  \href{http://dx.doi.org/10.1016/j.physrep.2007.04.003}{Phys. Rept. {\bf 445}
  (2007)  1--193},
\href{http://arxiv.org/abs/hep-th/0610327}{{\tt arXiv:hep-th/0610327}}.

\bibitem{Marchesano:2007de}
F.~Marchesano, \emph{{Progress in D-brane model building}},
  \href{http://dx.doi.org/10.1002/prop.200610381}{Fortsch. Phys. {\bf 55}
  (2007)  491--518},
\href{http://arxiv.org/abs/hep-th/0702094}{{\tt arXiv:hep-th/0702094}}.

\bibitem{Blumenhagen:2009qh}
R.~Blumenhagen, M.~Cvetic, S.~Kachru, and T.~Weigand, \emph{{{\small D}-Brane
  Instantons in Type {II} Orientifolds}},
  \href{http://dx.doi.org/10.1146/annurev.nucl.010909.083113}{Ann. Rev. Nucl.
  Part. Sci. {\bf 59} (2009)  269--296},
\href{http://arxiv.org/abs/0902.3251}{{\tt arXiv:0902.3251 [hep-th]}}.

\bibitem{Witten:1995gx}
E.~Witten, \emph{{Small Instantons in String Theory}},
  \href{http://dx.doi.org/10.1016/0550-3213(95)00625-7}{Nucl. Phys. {\bf B460}
  (1996)  541--559},
\href{http://arxiv.org/abs/hep-th/9511030}{{\tt arXiv:hep-th/9511030}}.

\bibitem{Douglas:1995bn}
M.~R. Douglas, \emph{{Branes within branes}},
\href{http://arxiv.org/abs/hep-th/9512077}{{\tt arXiv:hep-th/9512077}}.

\bibitem{Green:2000ke}
M.~B. Green and M.~Gutperle, \emph{{D-instanton induced interactions on a
  D3-brane}}, JHEP {\bf 02} (2000)  014,
\href{http://arxiv.org/abs/hep-th/0002011}{{\tt arXiv:hep-th/0002011}}.

\bibitem{Billo:2002hm}
M.~Billo, M.~Frau, I.~Pesando, F.~Fucito, A.~Lerda, and A.~Liccardo,
  \emph{{Classical gauge instantons from open strings}}, JHEP {\bf 02} (2003)
  045,
\href{http://arxiv.org/abs/hep-th/0211250}{{\tt arXiv:hep-th/0211250}}.

\bibitem{Blumenhagen:2006xt}
R.~Blumenhagen, M.~Cvetic, and T.~Weigand, \emph{{Spacetime instanton
  corrections in 4D string vacua - the seesaw mechanism for D-brane models}},
  \href{http://dx.doi.org/10.1016/j.nuclphysb.2007.02.016}{Nucl. Phys. {\bf
  B771} (2007)  113--142},
\href{http://arxiv.org/abs/hep-th/0609191}{{\tt arXiv:hep-th/0609191}}.

\bibitem{Ibanez:2006da}
L.~E. Ibanez and A.~M. Uranga, \emph{{Neutrino Majorana masses from string
  theory instanton effects}}, JHEP {\bf 03} (2007)  052,
\href{http://arxiv.org/abs/hep-th/0609213}{{\tt arXiv:hep-th/0609213}}.

\bibitem{Florea:2006si}
B.~Florea, S.~Kachru, J.~McGreevy, and N.~Saulina, \emph{{Stringy instantons
  and quiver gauge theories}}, JHEP {\bf 05} (2007)  024,
\href{http://arxiv.org/abs/hep-th/0610003}{{\tt arXiv:hep-th/0610003}}.

\bibitem{Donagi:2008ca}
R.~Donagi and M.~Wijnholt, \emph{{Model Building with F-Theory}},
\href{http://arxiv.org/abs/0802.2969}{{\tt arXiv:0802.2969 [hep-th]}}.

\bibitem{Beasley:2008dc}
C.~Beasley, J.~J. Heckman, and C.~Vafa, \emph{{GUTs and Exceptional Branes in
  F-theory - I}}, \href{http://dx.doi.org/10.1088/1126-6708/2009/01/058}{JHEP
  {\bf 01} (2009)  058},
\href{http://arxiv.org/abs/0802.3391}{{\tt arXiv:0802.3391 [hep-th]}}.

\bibitem{Beasley:2008kw}
C.~Beasley, J.~J. Heckman, and C.~Vafa, \emph{{GUTs and Exceptional Branes in
  F-theory - II: Experimental Predictions}},
  \href{http://dx.doi.org/10.1088/1126-6708/2009/01/059}{JHEP {\bf 01} (2009)
  059},
\href{http://arxiv.org/abs/0806.0102}{{\tt arXiv:0806.0102 [hep-th]}}.

\bibitem{Vafa:1996xn}
C.~Vafa, \emph{{Evidence for F-Theory}},
  \href{http://dx.doi.org/10.1016/0550-3213(96)00172-1}{Nucl. Phys. {\bf B469}
  (1996)  403--418},
\href{http://arxiv.org/abs/hep-th/9602022}{{\tt arXiv:hep-th/9602022}}.

\bibitem{Denef:2008wq}
F.~Denef, \emph{{Les Houches Lectures on Constructing String Vacua}},
\href{http://arxiv.org/abs/0803.1194}{{\tt arXiv:0803.1194 [hep-th]}}.

\bibitem{Heckman:2010bq}
J.~J. Heckman, \emph{{Particle Physics Implications of F-theory}},
\href{http://arxiv.org/abs/1001.0577}{{\tt arXiv:1001.0577 [hep-th]}}.

\bibitem{Berglund:2005dm}
  P.~Berglund and P.~Mayr,
  \emph{Non-perturbative superpotentials in F-theory and string duality},
\href{http://arxiv.org/abs/hep-th/0504058}{{\tt arXiv:hep-th/0504058}}.

\bibitem{Blumenhagen:2010ja}
R.~Blumenhagen, A.~Collinucci, and B.~Jurke, \emph{{On Instanton Effects in
  F-theory}},
\href{http://arxiv.org/abs/1002.1894}{{\tt arXiv:1002.1894 [hep-th]}}.

\bibitem{Cvetic:2010rq}
M.~Cvetic, I.~Garcia-Etxebarria, and J.~Halverson, \emph{{Global F-theory
  Models: Instantons and Gauge Dynamics}},
\href{http://arxiv.org/abs/1003.5337}{{\tt arXiv:1003.5337 [hep-th]}}.

\bibitem{Alim:2009bx}
M.~Alim, M.~Hecht, H.~Jockers, P.~Mayr, A.~Mertens and M.~Soroush, \emph{{Hints for Off-Shell Mirror Symmetry in type
  II/F-theory Compactifications}},
\href{http://arxiv.org/abs/0909.1842}{{\tt arXiv:0909.1842 [hep-th]}}.

\bibitem{Grimm:2009ef}
T.~W. Grimm, T.-W. Ha, A.~Klemm, and D.~Klevers, \emph{{Computing Brane and
  Flux Superpotentials in F-theory Compactifications}},
  \href{http://dx.doi.org/10.1007/JHEP04(2010)015}{JHEP {\bf 04} (2010)  015},
\href{http://arxiv.org/abs/0909.2025}{{\tt arXiv:0909.2025 [hep-th]}}.

\bibitem{Grimm:2009sy}
T.~W. Grimm, T.-W. Ha, A.~Klemm, and D.~Klevers, \emph{{Five-Brane
  Superpotentials and Heterotic/F-theory Duality}},
\href{http://arxiv.org/abs/0912.3250}{{\tt arXiv:0912.3250 [hep-th]}}.

\bibitem{Jockers:2009ti}
  H.~Jockers, P.~Mayr and J.~Walcher,
  \emph{On N=1 4d Effective Couplings for F-theory and Heterotic Vacua},
\href{http://arxiv.org/abs/0912.3265}{{\tt arXiv:0912.3265 [hep-th]}}.

\bibitem{Sen:1996vd}
A.~Sen, \emph{{F-theory and Orientifolds}},
  \href{http://dx.doi.org/10.1016/0550-3213(96)00347-1}{Nucl. Phys. {\bf B475}
  (1996)  562--578},
\href{http://arxiv.org/abs/hep-th/9605150}{{\tt arXiv:hep-th/9605150}}.

\bibitem{Seiberg:1994rs}
N.~Seiberg and E.~Witten, \emph{{Monopole Condensation, And Confinement In N=2
  Supersymmetric Yang-Mills Theory}},
  \href{http://dx.doi.org/10.1016/0550-3213(94)90124-4}{Nucl. Phys. {\bf B426}
  (1994)  19--52},
\href{http://arxiv.org/abs/hep-th/9407087}{{\tt arXiv:hep-th/9407087}}.

\bibitem{Seiberg:1994aj}
N.~Seiberg and E.~Witten, \emph{{Monopoles, duality and chiral symmetry
  breaking in N=2 supersymmetric QCD}},
  \href{http://dx.doi.org/10.1016/0550-3213(94)90214-3}{Nucl. Phys. {\bf B431}
  (1994)  484--550},
\href{http://arxiv.org/abs/hep-th/9408099}{{\tt arXiv:hep-th/9408099}}.

\bibitem{Banks:1996nj}
T.~Banks, M.~R. Douglas, and N.~Seiberg, \emph{{Probing F-theory with branes}},
  \href{http://dx.doi.org/10.1016/0370-2693(96)00808-8}{Phys. Lett. {\bf B387}
  (1996)  278--281},
\href{http://arxiv.org/abs/hep-th/9605199}{{\tt arXiv:hep-th/9605199}}.

\bibitem{Gava:1999ky}
E.~Gava, K.~S. Narain, and M.~H. Sarmadi, \emph{{Instantons in N = 2 Sp(N)
  superconformal gauge theories and the AdS/CFT correspondence}},
  \href{http://dx.doi.org/10.1016/S0550-3213(99)00751-8}{Nucl. Phys. {\bf B569}
  (2000)  183--208},
\href{http://arxiv.org/abs/hep-th/9908125}{{\tt arXiv:hep-th/9908125}}.

\bibitem{Billo:2006jm}
M.~Billo, M.~Frau, F.~Fucito, and A.~Lerda, \emph{{Instanton calculus in R-R
  background and the topological string}}, JHEP {\bf 11} (2006)  012,
\href{http://arxiv.org/abs/hep-th/0606013}{{\tt arXiv:hep-th/0606013}}.

\bibitem{Billo:2009di}
M.~Billo, L.~Ferro, M.~Frau, L.~Gallot, A.~Lerda, and I.~Pesando, \emph{{Exotic
  instanton counting and heterotic/type I' duality}},
  \href{http://dx.doi.org/10.1088/1126-6708/2009/07/092}{JHEP {\bf 07} (2009)
  092},
\href{http://arxiv.org/abs/0905.4586}{{\tt arXiv:0905.4586 [hep-th]}}.

\bibitem{Ito:2010vx}
  K.~Ito, H.~Nakajima, T.~Saka and S.~Sasaki,
  \emph{N=2 Instanton Effective Action in Omega-background and D3/D(-1)-brane
  System in R-R Background},
\href{http://arxiv.org/abs/1009.1212}{{\tt arXiv:1009.1212 [hep-th]}}.


\bibitem{Fucito:2009rs}
F.~Fucito, J.~F. Morales, and R.~Poghossian, \emph{{Exotic prepotentials from
  D(-1)D7 dynamics}},
  \href{http://dx.doi.org/10.1088/1126-6708/2009/10/041}{JHEP {\bf 10} (2009)
  041},
\href{http://arxiv.org/abs/0906.3802}{{\tt arXiv:0906.3802 [hep-th]}}.

\bibitem{Billo':2010bd}
M.~Billo, M.~Frau, F.~Fucito, A.~Lerda, J.~F. Morales, and R.~Poghossian,
  \emph{{Stringy instanton corrections to N=2 gauge couplings}},
  \href{http://dx.doi.org/10.1007/JHEP05(2010)107}{JHEP {\bf 05} (2010)  107},
\href{http://arxiv.org/abs/1002.4322}{{\tt arXiv:1002.4322 [hep-th]}}.

\bibitem{Nekrasov:2002qd}
N.~A. Nekrasov, \emph{{Seiberg-Witten Prepotential From Instanton Counting}},
  Adv. Theor. Math. Phys. {\bf 7} (2004)  831--864,
\href{http://arxiv.org/abs/hep-th/0206161}{{\tt arXiv:hep-th/0206161}}.

\bibitem{Flume:2002az}
R.~Flume and R.~Poghossian, \emph{{An algorithm for the microscopic evaluation
  of the coefficients of the Seiberg-Witten prepotential}},
  \href{http://dx.doi.org/10.1142/S0217751X03013685}{Int. J. Mod. Phys. {\bf
  A18} (2003)  2541},
\href{http://arxiv.org/abs/hep-th/0208176}{{\tt arXiv:hep-th/0208176}}.

\bibitem{Bruzzo:2002xf}
U.~Bruzzo, F.~Fucito, J.~F. Morales, and A.~Tanzini, \emph{{Multi-instanton
  calculus and equivariant cohomology}}, JHEP {\bf 05} (2003)  054,
\href{http://arxiv.org/abs/hep-th/0211108}{{\tt arXiv:hep-th/0211108}}.

\bibitem{Nekrasov:2003af}
N.~A. Nekrasov, \emph{{Seiberg-Witten prepotential from instanton counting}},
\href{http://arxiv.org/abs/hep-th/0306211}{{\tt arXiv:hep-th/0306211}}.

\bibitem{Nekrasov:2003rj}
N.~Nekrasov and A.~Okounkov, \emph{{Seiberg-Witten theory and random
  partitions}},
\href{http://arxiv.org/abs/hep-th/0306238}{{\tt arXiv:hep-th/0306238}}.

\bibitem{Bruzzo:2003rw}
U.~Bruzzo and F.~Fucito, \emph{{Superlocalization formulas and supersymmetric
  Yang-Mills theories}},
  \href{http://dx.doi.org/10.1016/j.nuclphysb.2003.11.033}{Nucl. Phys. {\bf
  B678} (2004)  638--655},
\href{http://arxiv.org/abs/math-ph/0310036}{{\tt arXiv:math-ph/0310036}}.

\bibitem{Lerche:1998nx}
W.~Lerche and S.~Stieberger, \emph{{Prepotential, mirror map and F-theory on
  K3}}, Adv. Theor. Math. Phys. {\bf 2} (1998)  1105--1140,
\href{http://arxiv.org/abs/hep-th/9804176}{{\tt arXiv:hep-th/9804176}}.

\bibitem{Lerche:1998gz}
W.~Lerche, S.~Stieberger, and N.~P. Warner, \emph{{Quartic gauge couplings from
  K3 geometry}}, Adv. Theor. Math. Phys. {\bf 3} (1999)  1575--1611,
\href{http://arxiv.org/abs/hep-th/9811228}{{\tt arXiv:hep-th/9811228}}.

\bibitem{Heckman:2010fh}
J.~J. Heckman and C.~Vafa, \emph{{An Exceptional Sector for F-theory GUTs}},
\href{http://arxiv.org/abs/1006.5459}{{\tt arXiv:1006.5459 [hep-th]}}.

\bibitem{Alday:2009aq}
L.~F. Alday, D.~Gaiotto, and Y.~Tachikawa, \emph{{Liouville Correlation
  Functions from Four-dimensional Gauge Theories}},
  \href{http://dx.doi.org/10.1007/s11005-010-0369-5}{Lett. Math. Phys. {\bf 91}
  (2010)  167--197},
\href{http://arxiv.org/abs/0906.3219}{{\tt arXiv:0906.3219 [hep-th]}}.

\bibitem{Zamolodchikov:1985ie}
A.~B. Zamolodchikov, \emph{{Conformal symmetry in two-dimensions: An explicit
  recurrence formula for the conformal partial wave amplitude}},
\href{http://dx.doi.org/10.1007/BF01214585}{Commun. Math. Phys. {\bf 96} (1984)
   419--422}.

\bibitem{Marshakov:2009gs}
A.~Marshakov, A.~Mironov, and A.~Morozov, \emph{{On Combinatorial Expansions of
  Conformal Blocks}},
\href{http://arxiv.org/abs/0907.3946}{{\tt arXiv:0907.3946 [hep-th]}}.

\bibitem{Marshakov:2009kj}
A.~Marshakov, A.~Mironov, and A.~Morozov, \emph{{Zamolodchikov asymptotic
  formula and instanton expansion in $\mathrm N=2$ SUSY $N_f=2N_c$ QCD}},
  \href{http://dx.doi.org/10.1088/1126-6708/2009/11/048}{JHEP {\bf 11} (2009)
  048},
\href{http://arxiv.org/abs/0909.3338}{{\tt arXiv:0909.3338 [hep-th]}}.

\bibitem{Poghossian:2009mk}
R.~Poghossian, \emph{{Recursion relations in CFT and $\mathcal N=2$ SYM
  theory}}, \href{http://dx.doi.org/10.1088/1126-6708/2009/12/038}{JHEP {\bf
  12} (2009)  038},
\href{http://arxiv.org/abs/0909.3412}{{\tt arXiv:0909.3412 [hep-th]}}.

\bibitem{Dorey:1996bf}
N.~Dorey, V.~V. Khoze, and M.~P. Mattis, \emph{{Multi-instanton calculus in N =
  2 supersymmetric gauge theory. II: Coupling to matter}},
  \href{http://dx.doi.org/10.1103/PhysRevD.54.7832}{Phys. Rev. {\bf D54} (1996)
   7832--7848},
\href{http://arxiv.org/abs/hep-th/9607202}{{\tt arXiv:hep-th/9607202}}.

\bibitem{Dorey:1996bn}
N.~Dorey, V.~V. Khoze, and M.~P. Mattis, \emph{{On N = 2 supersymmetric {QCD}
  with 4 flavors}},
  \href{http://dx.doi.org/10.1016/S0550-3213(97)00132-6}{Nucl. Phys. {\bf B492}
  (1997)  607--622},
\href{http://arxiv.org/abs/hep-th/9611016}{{\tt arXiv:hep-th/9611016}}.

\bibitem{Billo':2009gc}
M.~Billo, M.~Frau, L.~Gallot, A.~Lerda, and I.~Pesando, \emph{{Classical
  solutions for exotic instantons?}},
  \href{http://dx.doi.org/10.1088/1126-6708/2009/03/056}{JHEP {\bf 03} (2009)
  056},
\href{http://arxiv.org/abs/0901.1666}{{\tt arXiv:0901.1666 [hep-th]}}.

\bibitem{Moore:1998et}
G.~W. Moore, N.~Nekrasov, and S.~Shatashvili, \emph{{D-particle bound states
  and generalized instantons}},
  \href{http://dx.doi.org/10.1007/s002200050016}{Commun. Math. Phys. {\bf 209}
  (2000)  77--95},
\href{http://arxiv.org/abs/hep-th/9803265}{{\tt arXiv:hep-th/9803265}}.

\bibitem{Gutperle:1999xu}
M.~Gutperle, \emph{{Heterotic/type I duality, D-instantons and a N = 2 AdS/CFT
  correspondence}}, \href{http://dx.doi.org/10.1103/PhysRevD.60.126001}{Phys.
  Rev. {\bf D60} (1999)  126001},
\href{http://arxiv.org/abs/hep-th/9905173}{{\tt arXiv:hep-th/9905173}}.

\bibitem{Kiritsis:2000zi}
E.~Kiritsis, N.~A. Obers, and B.~Pioline, \emph{{Heterotic/type II triality and
  instantons on K3}}, JHEP {\bf 01} (2000)  029,
\href{http://arxiv.org/abs/hep-th/0001083}{{\tt arXiv:hep-th/0001083}}.

\bibitem{Green:1987mn}
M.~B. Green, J.~H. Schwarz, and E.~Witten, \emph{{Superstring Theory. Vol. 2:
  Loop amplitudes, Anomalies and Phenomenology}},. Cambridge Univ. Pr. ( 1987)
  ( Cambridge Monographs On Mathematical Physics).

\bibitem{D'Hoker:1999ft}
E.~D'Hoker and D.~H. Phong, \emph{{Lectures on supersymmetric Yang-Mills theory
  and integrable systems}},
\href{http://arxiv.org/abs/hep-th/9912271}{{\tt arXiv:hep-th/9912271}}.

\bibitem{privcomm}
F.~Fucito, J.~F. Morales, and R.~Poghossian, \emph{{private communication}}.

\bibitem{Matone:1995rx}
M.~Matone, \emph{{Instantons and recursion relations in N=2 SUSY gauge
  theory}}, \href{http://dx.doi.org/10.1016/0370-2693(95)00920-G}{Phys. Lett.
  {\bf B357} (1995)  342--348},
\href{http://arxiv.org/abs/hep-th/9506102}{{\tt arXiv:hep-th/9506102}}.

\bibitem{Grimm:2007tm}
T.~W. Grimm, A.~Klemm, M.~Marino, and M.~Weiss, \emph{{Direct integration of
  the topological string}}, JHEP {\bf 08} (2007)  058,
\href{http://arxiv.org/abs/hep-th/0702187}{{\tt arXiv:hep-th/0702187}}.

\end{thebibliography}
\end{document}